%% file: Kampen.Salazar.ea.ECC22.tex
\documentclass[conference]{ieeeconf}
\renewcommand{\baselinestretch}{0.98}
\IEEEoverridecommandlockouts

\input{Report/Packages.tex}
\input{Report/Acronyms.tex}

\input{Report/def.tex}
\newcommand{\pushright}[1]{\ifmeasuring@#1\else\omit\hfill$\displaystyle#1$\fi\ignorespaces}
\newcommand{\pushleft}[1]{\ifmeasuring@#1\else\omit$\displaystyle#1$\hfill\fi\ignorespaces}
\makeatother
\newif\ifmargincomments 
\margincommentstrue

\newif\ifextendedversion 
\extendedversionfalse

\ifmargincomments

\else

\fi
\begin{document}

\title{\bf Maximum-distance Race Strategies for a\\Fully Electric Endurance Race Car
}

\author{Jorn van Kampen, Thomas Herrmann, and Mauro Salazar%
	\thanks{Jorn van Kampen and Mauro Salazar are with the Control Systems Technology section, Department of Mechanical Engineering, Eindhoven University of Technology (TU/e), Eindhoven, 5600 MB, The Netherlands.
	E-mails: {\tt\footnotesize j.h.e.v.kampen@student.tue.nl}, {\tt\footnotesize m.r.u.salazar@tue.nl} \newline Thomas Herrmann is with the Institute of Automotive Technology, Department of Mechanical Engineering, Technical University of Munich (TUM).
	E-mail: {\tt\footnotesize thomas.herrmann@tum.de}}
}


\maketitle
\begin{abstract}
This paper presents a bi-level optimization framework to compute the maximum-distance stint and charging strategies for a fully electric endurance race car.
Thereby, the lower level computes the minimum-stint-time Powertrain Operation (PO) for a given battery energy budget and stint length, whilst the upper level leverages that information to jointly optimize the stint length, charge time and number of pit stops, in order to maximize the driven distance in the course of a fixed-time endurance race.
Specifically, we first extend a convex lap time optimization framework to capture multiple laps and force-based electric motor models, and use it to create a map linking the charge time and stint length to the achievable stint time.
Second, we leverage the map to frame the maximum-race-distance problem as a mixed-integer second order conic program that can be efficiently solved to the global optimum with off-the-shelf optimization algorithms.
Finally, we showcase our framework on a \unit[6]{h} race around the Zandvoort circuit.
Our results show that a flat-out strategy can be extremely detrimental, and that, compared to when the stints are optimized for a fixed number of pit stops, jointly optimizing the stints and number of pit stops can increase the driven distance of several laps.
\end{abstract}

\input{Sections/Introduction}
\input{Sections/Methods}
\input{Sections/Results}
\input{Sections/Conclusion}

\section*{Acknowledgment}
\noindent We thank Dr.~I.~New, Ir.~S.~Broere and Ir.~M.~Konda for proofreading, and G.~Delissen for the photograph. This paper was partly supported by the NEON research project (project number 17628 of the Crossover program which is (partly) financed by the Dutch Research Council (NWO)).

\bibliographystyle{IEEEtran}
\bibliography{Report/bibliography.bib,Report/main.bib,Report/SML_papers.bib}\newpage

\end{document}

%% file: Report/Packages.tex
\usepackage{cite}
\usepackage[acronym]{glossaries}
\usepackage{amsmath,amssymb,amsfonts}
\usepackage{algorithmic}
\usepackage{graphicx}
\usepackage{xcolor}
\def\BibTeX{{\rm B\kern-.05em{\sc i\kern-.025em b}\kern-.08em
    T\kern-.1667em\lower.7ex\hbox{E}\kern-.125emX}}

\usepackage{units}
\usepackage{mathtools} 
\usepackage[colorlinks=true, allcolors=blue]{hyperref}
\usepackage{siunitx}

\usepackage[hyphenbreaks]{breakurl}
\usepackage{hyperref}
\usepackage{amsthm}
\newcounter{thm}
\newtheorem{prob}[thm]{Problem}
\usepackage{lettrine}

\usepackage{scrextend} 

\usepackage{tabularx}
\usepackage[english]{babel}
\usepackage{cite}

\usepackage{array}
\usepackage{color}
\usepackage{amssymb}
\usepackage{multicol}

\DeclareUnicodeCharacter{2212}{-}

\usepackage{todonotes}

\usepackage{tikz}
\usetikzlibrary{arrows,calc}
\usepackage{pgfplots}
\usepgfplotslibrary{groupplots}

%% file: Report/Acronyms.tex
\newacronym{acr:cvt}{CVT}{continuously variable transmission}
\newacronym{acr:dp}{DP}{dynamic programming}
\newacronym{acr:ecms}{ECMS}{equivalent consumption minimization strategies}
\newacronym{acr:eltms}{ELTMS}{equivalent lap time minimization strategies}
\newacronym{acr:em}{EM}{electric motor}
\newacronym{acr:es2k}{ES2K}{Energy Storage to Kinetic}
\newacronym{acr:F1}{F1}{Formula 1}
\newacronym{acr:FIA}{FIA}{F\'{e}d\'{e}ration Internationale de l'Automobile}
\newacronym{acr:fgt}{FGT}{fixed-gear transmission}
\newacronym{acr:FD}{FD}{final drive}
\newacronym{acr:ice}{ICE}{internal combustion engine}
\newacronym{acr:k2es}{K2ES}{Kinetic to Energy Storage}
\newacronym{acr:mgu}{MGU}{motor generator unit}
\newacronym{acr:mguh}{MGU-H}{motor generator unit heat}
\newacronym{acr:mguk}{MGU-K}{motor generator unit kinetic}
\newacronym{acr:mpc}{MPC}{model predictive control}
\newacronym[description={Energy Management Strategy}, \glslongpluralkey={Energy Management Strategies},\glsshortpluralkey={EMSs}]{EMS}{EMS}{Energy Management Strategy}%
\newacronym{acr:pmp}{PMP}{Pontryagin's Minimum Principle}
\newacronym{acr:pu}{PU}{power unit}
\newacronym[description={Powertrain Operation}, \glslongpluralkey={Powertrain Operations},\glsshortpluralkey={POs}]{acr:PO}{PO}{Powertrain Operation}%
\newacronym{acr:rmse}{RMSE}{root-mean-square error}
\newacronym{acr:socp}{SOCP}{second-order cone program}
\newacronym{acr:soe}{SoE}{State of Energy}


%% file: Report/def.tex
\newcommand{\cS}{\mathcal{S}}

\newcommand{\sN}{\mathbb{N}}

\newcommand{\dtds}{\frac{\textnormal{d}t}{\textnormal{d}s}}

%% file: Sections/Introduction.tex
\section{Introduction}

\lettrine{T}{he electrification} of race cars has been increasing in popularity over the last years, with the advent of hybrid electric Formula 1 cars and Le Mans Hypercars, and battery electric vehicles in Formula E.
In a setting where every millisecond counts, it is of paramount importance to profit the most of the energy stored on-board via optimized \gls{EMS}.
In this context, the possibility of recharging the battery in the course of the race further complicates the problem, requiring race engineers to strike the best trade-off between reducing consumptions and pit-stops at the cost of lap-time, or driving faster with more pit-stops.
This conflict becomes particularly imminent in endurance racing, where the objective is to maximize the driven distance in a fixed amount of time, which can range up to \SI{24}{\hour}~\cite{FIA2021}.
In this setting, the car has to be strategically recharged during pit stops in order to maintain a competitive performance, maximizing the distance driven.
This calls for algorithms to compute the maximum-distance race strategies that provide the number of pit stops during the race, the number of laps driven per stint ( referred to as stint lengths) and charge time (which is directly correlated to charge energy), whilst accounting for the optimal energy management strategies and \gls{acr:PO}.
Against this backdrop, this paper presents a bi-level optimization framework to compute the maximum-distance race strategies with global optimality guarantees.


\begin{figure}
	\centering
	\includegraphics[width=1\linewidth,trim= 0 150 0 250,clip]{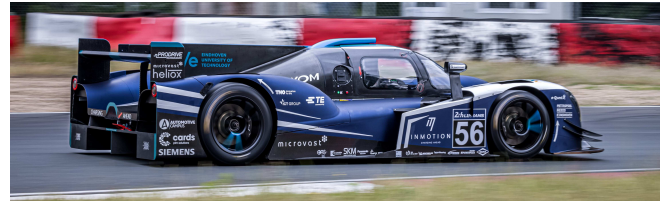}
	\caption{InMotion's fully electric endurance race car.}
	\label{fig:inmotion-car}
\end{figure}

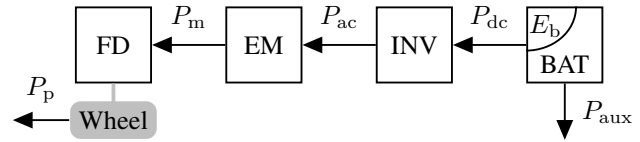
\begin{figure}[!t]
	\centering
	\input{./Figures/IM_topology_no_temp.tex}
	\caption{Schematic layout of the electric race car powertrain topology consisting of a battery (BAT), inverter (INV), electric machine (EM) and final drive (FD). The arrows indicate positive power flows.}
	\label{fig:topology}
\end{figure}

\subsubsection*{Related Literature}
This work pertains to two main research streams: single-lap optimization of the \glspl{EMS} jointly with the vehicle trajectory or for a given race line, and full-race optimization via simulations.

Several authors optimized the minimum-lap-time race line for a single race lap using both direct and indirect optimization methods~\cite{LotEvangelou2013, SedlacekOdenthalEtAl2020, Casanova2000, LimebeerPerantoni2014, ChristWischnewskiEtAl2019, DalBiancoLotEtAl2017, HeilmeierWischnewskiEtAl2020}. Some of these studies also include a maximum energy consumption per lap to approach racing conditions~\cite{HerrmannChristEtAl2019}. Similar approaches extend the minimum-lap-time problems to minimum-race-time problems. They consider temperature dynamics, and optimize for multiple consecutive race laps to enable a variable amount of energy consumed per lap, but formulate the optimization problem in space domain for an a priori-known number of laps~\cite{HerrmannPassigatoEtAl2020,LiuFotouhiEtAl2020}. Finally, considering the race line to be fixed, multi-lap \glspl{EMS} are optimized, leveraging nonlinear optimization techniques~\cite{HerrmannSauerbeckEtAl2021} or artificial neural networks~\cite{LiuFotouhi2020}. However, these papers lack global optimality guarantees.

Against this backdrop, assuming the race line to be available in the form of a maximum speed profile, convex optimization has been successfully leveraged to compute the globally optimal \glspl{EMS} for hybrid and fully electric race vehicles~\cite{EbbesenSalazarEtAl2018,SalazarElbertEtAl2017}, also including gear shift strategies~\cite{DuhrChristodoulouEtAl2020}, different transmission technologies~\cite{BorsboomFahdzyanaEtAl2021} and thermal limitations~\cite{LocatelloKondaEtAl2020}. Yet these methods are focused on single-lap problems and do not capture pit-stops and recharging processes.

A final relevant research stream involves race simulations, in which entire races are optimized on a per lap basis~\cite{HeilmeierGrafEtAl2018, WestLimebeer2020}. However, these studies mainly focus on optimal tire strategies by modeling their degradation as a lap time increase and do not capture the charging and \gls{acr:PO} strategies.
In conclusion, to the best of the authors' knowledge, there are no methods specifically focusing on race strategies in endurance scenarios, whereby the \gls{acr:PO} within a stint and the stints themselves are jointly optimized.

\subsubsection*{Statement of Contributions}
This paper presents a bi-level mixed-integer convex optimization framework to efficiently compute the globally optimal, maximum-distance endurance race strategies and the corresponding \gls{acr:PO} in the individual stints.
Our low-level algorithm computes the optimal stint time for a given number of laps and different levels of recharged battery energy. To preserve convexity, we describe the \gls{acr:em} efficiency by using speed-dependent in- and output forces. Subsequently, we fit the relationship between the stint length, the charged energy, and the achievable stint time as a second-order conic constraint, which we leverage in the high-level algorithm.
Thereby we frame the maximum-distance race problem as a mixed-integer second-order conic program which jointly optimizes the stint length, the charge time---i.e., the charge energy---and the number of pit stops.
The resulting problem can be rapidly solved with off-the-shelf numerical solvers with global optimality guarantees.
Finally, we showcase our framework on the Zandvoort circuit for the vehicle shown in Fig.~\ref{fig:inmotion-car}, highlighting the importance of jointly optimizing the number of pit stops with the stint lengths and charging strategies.

\subsubsection*{Organization}
The remainder of this paper is structured as follows: Section~\ref{sec:low level} presents the minimum-stint-time control problem, after which Section~\ref{sec:high level} frames the maximum-race-distance control problem. We showcase our framework for a \unit[6]{h} race in Section~\ref{Results}. Finally, Section~\ref{Conclusion} draws the conclusions and provides an outlook on future research.

%% file: Figures/IM_topology_no_temp.tex
\tikzstyle{simpleNode} = [rectangle, minimum width=1cm, minimum height=1cm,text centered, draw=black, line width=0.25mm]
\tikzstyle{arrow} = [->,-triangle 45,line width=0.25mm,black]
\tikzstyle{lineE} = [-,line width=0.5mm,gray!50,solid]
\tikzstyle{lineM} = [-,line width=0.25mm,black]
\tikzset{
	relative at/.style n args = {3}{
		at = {({$(#1.west)!#2!(#1.east)$} |- {$(#1.south)!#3!(#1.north)$})}
	}
}
\begin{tikzpicture}[node distance=2cm]
\coordinate (Zero) at (0,0);
\node (FD) [simpleNode, align=center] {FD};
\node (EM) [simpleNode, align=center, right of=FD] {EM};
\node (INV) [simpleNode, align=center, right of=EM] {INV};
\node (BAT) [simpleNode, align=center, right of=INV] {};
\node[relative at={BAT}{0.5}{0.25}] {BAT};
\node[relative at={BAT}{0.25}{0.75}] {$E_\mathrm{b}$};
\node (Wheel) [simpleNode, align=center, below of=FD, yshift=1cm, minimum height=0.5cm, draw=none, fill=gray!50, rounded corners] {Wheel};
\draw [lineM] ($(BAT.west)+(0,-0.075)$) to [out=0,in=270] ($(BAT.north)+(0.15,0)$);
\draw [lineE] (FD.south) -- node[right] {} ($(Wheel.north)-(0,0.1)$);
\draw [arrow] (Wheel.west) -- node[above, yshift=0.1cm] {$P_\mathrm{p}$} ($(Wheel.west) + (-0.75,0)$);
\draw [arrow] (EM.west) -- node[above, yshift=0.1cm] {$P_\mathrm{m}$} (FD.east);
\draw [arrow] (INV.west) -- node[above, yshift=0.1cm] {$P_\mathrm{ac}$} (EM.east);
\draw [arrow] (BAT.west) -- node[above, yshift=0.1cm] {$P_\mathrm{dc}$} (INV.east);
\draw [arrow] (BAT.south) -- node[right, xshift=0.1cm] {$P_\mathrm{aux}$} ($(BAT.south) - (0,0.75)$);
\end{tikzpicture}

%% file: Sections/Methods.tex
\section{Low-level Stint Optimization} \label{sec:low level}
This section illustrates the minimum-stint-time control problem in space domain, since minimizing the stint time given a fixed distance represents the dual problem of maximizing distance within a fixed time period. We extend an existing convex framework~\cite{BorsboomFahdzyanaEtAl2021} to allow multi-lap optimization, whilst improving the \gls{acr:em} model accuracy by considering a pre-defined fixed-gear transmission ratio. Thereby we separate the \gls{acr:em} and inverter model to allow future extensions to temperature models. From the time-optimal control problem, we obtain the minimum stint time for a given stint length and available battery energy (which can be equivalently expressed in terms of charging time). 

Fig.~\ref{fig:topology} shows a schematic representation of the powertrain topology of the electric race car. The EM propels the rear wheels through a fixed \gls{acr:FD}, while receiving energy from the battery pack via the inverter. As with most electric vehicles, the EM can also operate as a generator, thus we account for a bi-directional energy flow between the battery and the wheels. In addition, we consider auxiliary components that are powered from the main battery as a uni-directional energy flow.

In reality, the driver controls the EM torque through the accelerator pedal and as such we define the mechanical EM power $P_{\mathrm{m}}$ as the input variable. As state variables, we choose the battery energy $E_{\mathrm{b}}$ and the kinetic energy of the vehicle $E_{\mathrm{kin}}$. The remaining energy flows between the powertrain components are the propulsion power $P_{\mathrm{p}}$, electrical EM power $P_{\mathrm{ac}}$, electrical inverter power $P_{\mathrm{dc}}$ and auxiliary supply $P_{\mathrm{aux}}$. Since we formulate the control problem in space domain, we ultimately define the model in terms of forces rather than power. Thus we divide power by the vehicle velocity, since the space-derivative of energy is defined with respect to the vehicle.

\subsection{Objective and Longitudinal Dynamics}


In racing, the objective is to minimize the lap times over the entire race. Since we only consider a stint in the low-level control problem, the objective is to minimize the stint time $t_{\mathrm{stint}}$, which is defined as
\par\nobreak\vspace{-5pt}
\begingroup
\allowdisplaybreaks
\begin{small}
	\begin{equation}
		\min t_{\mathrm{stint}} = \min \int_{0}^{S_{\mathrm{stint}}} \frac{\mathrm{d}t}{\mathrm{d}s}(s) \ \mathrm{d}s,
		\label{eq:low_obj}
	\end{equation}
\end{small}%
\endgroup
where $S_{\mathrm{stint}}$ is the stint length in terms of distance and $\frac{\mathrm{d}t}{\mathrm{d}s}(s)$ is the lethargy, which is the inverse of the vehicle velocity $v(s)\geq0$. To implement the lethargy as a convex constraint, we define
\par\nobreak\vspace{-5pt}
\begingroup
\allowdisplaybreaks
\begin{small}
	\begin{equation}
		\frac{\mathrm{d}t}{\mathrm{d}s}(s) \geq \frac{1}{v(s)},
		\label{eq:lethargy}
	\end{equation}
\end{small}%
\endgroup
which is a convex relaxation that holds with equality in case of an optimal solution~\cite{EbbesenSalazarEtAl2018}.

Since the goal of this paper is to study the optimal race strategy and \gls{acr:PO} rather than studying the effect of vehicle dynamics, we model the vehicle as a point mass, for which the longitudinal dynamics are written as 

\par\nobreak\vspace{-5pt}
\begingroup
\allowdisplaybreaks
\begin{small}
	\begin{equation}
		\frac{\mathrm{d}}{\mathrm{d}s}E_{\mathrm{kin}}(s) = F_{\mathrm{p}}(s)-F_{\mathrm{d}}(s)-F_{\mathrm{brake}}(s),
		\label{eq:longitudinal}
	\end{equation}
\end{small}%
\endgroup
where $F_{\mathrm{p}}(s)$ is the propulsion force, $F_{\mathrm{d}}(s)$ is the drag force and $F_{\mathrm{brake}}(s)$ is the force from the mechanical brakes. The drag force is defined as the sum of the aerodynamic drag, the rolling resistance and the gravitational force as
\par\nobreak\vspace{-5pt}
\begingroup
\allowdisplaybreaks
\begin{small}
	\begin{multline} \label{eq:F_rl}
		F_{\mathrm{d}}(s) = \frac{c_{\mathrm{d}} \cdot A_{\mathrm{f}} \cdot \rho}{m_{\mathrm{tot}}} \cdot E_{\mathrm{kin}}(s) +  c_{\mathrm{r}}\cdot(m_{\mathrm{tot}}\cdot g\cdot \cos(\theta(s)) + \\ F_{\mathrm{down}}(s)) + m_{\mathrm{tot}}\cdot g\cdot \sin(\theta(s)),
	\end{multline}	
\end{small}%
\endgroup
where $m_{\mathrm{tot}}$ is the total mass of the vehicle, $c_{\mathrm{d}}$ is the air drag coefficient, $A_{\mathrm{f}}$ is the frontal area of the vehicle, $\rho$ is the air density, $c_{\mathrm{r}}$ is the rolling resistance coefficient, $g$ is the gravitational constant, $\theta(s)$ is the inclination of the track and $F_{\mathrm{down}}(s)$ is the aerodynamic downforce defined by
\par\nobreak\vspace{-5pt}
\begingroup
\allowdisplaybreaks
\begin{small}
	\begin{equation}\label{eq:F_down}
		F_{\mathrm{down}}(s) = \frac{c_{\mathrm{l}} \cdot A_{\mathrm{f}} \cdot \rho}{m_{\mathrm{tot}}}\cdot E_{\mathrm{kin}}(s),
	\end{equation}
\end{small}%
\endgroup
where $c_{\mathrm{l}}$ is the aerodynamic lift coefficient. To account for the losses in the final drive under bi-directional power flow, we write \eqref{eq:longitudinal} as two inequality constraints according to
%

\par\nobreak\vspace{-5pt}
\begingroup
\allowdisplaybreaks
\begin{small}
	\begin{equation} \label{eq:force_balance_trac}
		\frac{\mathrm{d}}{\mathrm{d}s}E_{\mathrm{kin}}(s) \leq F_{\mathrm{m}}(s)\cdot \eta_{\mathrm{fd}} -F_{\mathrm{d}}(s)-F_{\mathrm{brake}}(s),
	\end{equation} 
\end{small}%
\endgroup
\par\nobreak\vspace{-5pt}
\begingroup
\allowdisplaybreaks
\begin{small}
	\begin{equation} \label{eq:force_balance_reg}
		\frac{\mathrm{d}}{\mathrm{d}s}E_{\mathrm{kin}}(s) \leq F_{\mathrm{m}}(s)\cdot \frac{1}{\eta_{\mathrm{fd}}} -F_{\mathrm{d}}(s)-F_{\mathrm{brake}}(s),
	\end{equation}
\end{small}%
\endgroup
where $F_{\mathrm{m}}(s)$ is the mechanical output force from the EM and $\eta_{\mathrm{fd}}$ is the efficiency of the final drive, assumed constant. Due to the objective~\eqref{eq:low_obj}, in case of traction, \eqref{eq:force_balance_trac} will hold with equality, whilst in case of regenerative braking, \eqref{eq:force_balance_reg} will hold with equality, thus capturing the bi-directional power flow.
The relation between the kinetic energy and velocity of the vehicle is defined by a convex relaxation as
\par\nobreak\vspace{-5pt}
\begingroup
\allowdisplaybreaks
\begin{small}
	\begin{equation} \label{eq:E_kin}
		\frac{1}{2} \cdot m_{\mathrm{tot}}\cdot v^2(s) \leq	E_{\mathrm{kin}}(s) \leq \frac{1}{2} \cdot m_{\mathrm{tot}}\cdot v_{\mathrm{max}}^2(s),
	\end{equation}
\end{small}%
\endgroup
where $v_{\mathrm{max}}(s)$ is the maximum velocity possible without exceeding the tire grip limitations on the race track. This maximum velocity profile can be pre-computed according to the method shown in~\cite{BorsboomFahdzyanaEtAl2021}.

In contrast to single-lap scenarios, a stint is represented by the vehicle starting and stopping at the pit box with a certain number of flying laps in between. However, since we are working in space domain, the lethargy would diverge to infinity for zero velocity. To solve this issue, we define a minimal velocity $v_{\mathrm{min}}$ close to standstill and enforce this value to the initial and final velocity with
\par\nobreak\vspace{-5pt}
\begingroup
\allowdisplaybreaks
\begin{small}
	\begin{equation} \label{eq:E0_ES}
		E_{\mathrm{kin}}(0) = E_{\mathrm{kin}}(S_{\mathrm{stint}}) = \frac{1}{2}\cdot m_{\mathrm{tot}}\cdot v_{\mathrm{min}}^2.
	\end{equation}
\end{small}%
\endgroup
When driving through the pit lane, the vehicle should adhere to a strict speed limit, of which the exact value is track-dependent. Therefore, we define an upper bound $v_{\mathrm{pit,max}}$ on the vehicle velocity when the vehicle is exiting or entering the pit as
\par\nobreak\vspace{-5pt}
\begingroup
\allowdisplaybreaks
\begin{small}
	\begin{equation} \label{eq:Ek_pit}
		E_{\mathrm{kin}}(s) \leq \frac{1}{2}\cdot m_{\mathrm{tot}}\cdot v_{\mathrm{pit,max}}^2 \quad \forall s \in \cS_{\mathrm{pit}},
	\end{equation}
\end{small}%
\endgroup
where $\cS_{\mathrm{pit}}$ is the set of distance-based positions that are part of the pit lane.
Finally, we have to consider the maximum deceleration of the vehicle whenever the maximum velocity profile is not an active constraint, e.g., during braking before the pit entry. Assuming straight line braking, we can express the maximum deceleration as a lower bound on the kinetic energy with
\par\nobreak\vspace{-5pt}
\begingroup
\allowdisplaybreaks
\begin{small}
	\begin{equation} \label{eq:max_brake}
		\frac{\mathrm{d}E_{\mathrm{kin}}}{\mathrm{d}s}(s) \geq -F_{\mathrm{d}}(s)-\mu\cdot(m_{\mathrm{tot}}\cdot g + F_{\mathrm{down}}(s)),
	\end{equation}
\end{small}%
\endgroup
where $\mu$ is the friction coefficient of the tires. 


\subsection{Electric Machine}\label{sec:EM}
This section derives a convex representation of the operating limits and power losses of the \gls{acr:em}. In general, we can distinguish between a maximum torque and maximum power operating region for an \gls{acr:em}. Translating this to constraints in space domain results in a lower and upper bound on the mechanical output force of the EM for the maximum torque region as
\par\nobreak\vspace{-5pt}
\begingroup
\allowdisplaybreaks
\begin{small}
	\begin{equation} \label{eq:Tmax}
		F_{\mathrm{m}}(s) \in \left[-\frac{T_{\mathrm{m,max}}\cdot \gamma_{\mathrm{fd}}}{r_{\mathrm{w}}},\frac{T_{\mathrm{m,max}}\cdot \gamma_{\mathrm{fd}}}{r_{\mathrm{w}}}\right],
	\end{equation}
\end{small}%
\endgroup
where $T_{\mathrm{m,max}}$ is the maximum torque the \gls{acr:em} can deliver, $\gamma_{\mathrm{fd}}$ is the final drive ratio and $r_{\mathrm{w}}$ is the radius of the rear wheels. Note that we include the final drive ratio, as we define the space-derivatives with respect to the vehicle reference frame. Similarly, the mechanical output force of the \gls{acr:em} within the maximum power region is bounded as
\par\nobreak\vspace{-5pt}
\begingroup
\allowdisplaybreaks
\begin{small}
	\begin{equation} 
		F_{\mathrm{m}}(s) \in \left[-P_{\mathrm{m,max}}\cdot \frac{\mathrm{d}t}{\mathrm{d}s}(s), P_{\mathrm{m,max}}\cdot \frac{\mathrm{d}t}{\mathrm{d}s}(s) \right],
		\label{eq:P_em,max}
	\end{equation}
\end{small}%
\endgroup
where $P_{\mathrm{m,max}}$ is the maximum power the \gls{acr:em} can deliver. 

We model the \gls{acr:em} \emph{force} losses $F_{\mathrm{m,loss}}(s)$ rather than the power losses as a function of the vehicle velocity and force of the EM. In general, an \gls{acr:em} efficiency map shows large losses at low rotational velocities. Therefore, we want to include a term in our losses fit that is inversely proportional to the vehicle velocity. To ensure convexity, we model the \gls{acr:em} losses as
\par
\begin{figure}
 	\vspace{4mm}
	\centering
	\includegraphics[width=\linewidth,trim=0 0 0 10]{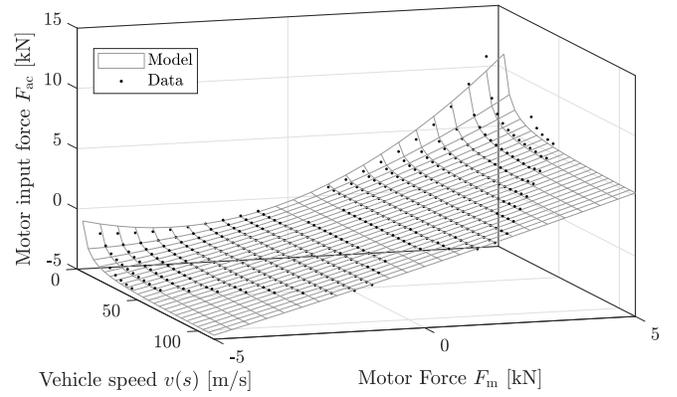}
	\caption{A speed- and torque-dependent model of the EM. The normalized root-mean-square error (RMSE) of the model is 1.49\% w.r.t.\ the maximum motor input force $F_\mathrm{ac}$.  }
	\label{fig:emlossfit}
\end{figure}

\nobreak\vspace{-5pt}
\begingroup
\allowdisplaybreaks
\begin{small}
	\begin{equation}
		F_{\mathrm{m,loss}}(s) = x_{\mathrm{m}}^{\top}(s) Q_{\mathrm{m}}x_{\mathrm{m}}(s),
	\end{equation}
\end{small}%
\endgroup
where $x_{\mathrm{m}}(s) = \left[\frac{1}{\sqrt{v(s)}} \  \sqrt{v(s)} \  \frac{F_{\mathrm{m}}(s)}{\sqrt{v(s)}}\right]^\top$ and $Q_{\mathrm{m}}$ is a symmetric positive semi-definite matrix of coefficients, of which the values are determined through semi-definite programming. Fig.~\ref{fig:emlossfit} shows the \gls{acr:em} input force as a function of the \gls{acr:em} output force and vehicle speed for the convex model and for the reference data.
To implement the losses in a convex manner, we take the relation of the electrical \gls{acr:em} input force $F_{\mathrm{ac}}(s)$ to the mechanical output force as
\par\nobreak\vspace{-5pt}
\begingroup
\allowdisplaybreaks
\begin{small}
	\begin{equation}
		F_{\mathrm{ac}}(s) = F_{\mathrm{m}}(s) + F_{\mathrm{m,loss}}(s),
	\end{equation}
\end{small}%
\endgroup
substitute the loss model, relax it and rewrite to a convex relaxation as
\par\nobreak\vspace{-5pt}
\begingroup
\allowdisplaybreaks
\begin{small}
	\begin{equation} \label{eq:eff_EM}
		(F_{\mathrm{ac}}(s)-F_{\mathrm{m}}(s))\cdot v(s) \geq y_{\mathrm{m}}^{\top}(s) Q_{\mathrm{m}}y_{\mathrm{m}}(s),
	\end{equation}
\end{small}%
\endgroup
where $y_{\mathrm{m}}(s) = \left[1 \ v(s) \  F_{\mathrm{m}}(s)\right]^\top$.
\par

\subsection{Inverter}
In this section, we derive a quadratic model for the inverter losses. We apply the general quadratic power loss model of the form
\par\nobreak\vspace{-5pt}
\begingroup
\allowdisplaybreaks
\begin{small}
	\begin{equation}
		P_{\mathrm{dc}}(s) = \alpha \cdot P_{\mathrm{ac}}^2(s)+P_{\mathrm{ac}}(s),
	\end{equation}
\end{small}%
\endgroup
where $\alpha$ is an efficiency parameter, subject to identification. Converting this constraint to forces, rewriting and relaxing results in
\par\nobreak\vspace{-5pt}
\begingroup
\allowdisplaybreaks
\begin{small}
	\begin{equation} \label{eq:eff_inverter}
		(F_{\mathrm{dc}}(s)-F_{\mathrm{ac}}(s))\cdot \frac{\mathrm{d}t}{\mathrm{d}s}(s) \geq \alpha \cdot F_{\mathrm{ac}}^2(s),
	\end{equation}
\end{small}%
\endgroup
where $F_{\mathrm{dc}}(s)$ is the force equivalent to the electrical inverter input power. 
\par
\subsection{Battery}\label{sec:batt}
This section derives a model for the battery efficiency and the power-split between the electrical inverter power and the auxiliary component power. The latter can be observed from Fig.~\ref{fig:topology} and is written as
\par\nobreak\vspace{-5pt}
\begingroup
\allowdisplaybreaks
\begin{small}
	\begin{equation}
		P_{\mathrm{b}}(s) = P_{\mathrm{dc}}(s) + P_{\mathrm{aux}},
		\label{eq:P_bat_split}
	\end{equation}
\end{small}%
\endgroup  
where $P_{\mathrm{b}}(s)$ is the battery power at the terminals. Here, the auxiliary component supply is assumed to be constant and uni-directional, while the other powers are bi-directional. Converting \eqref{eq:P_bat_split} to forces results in
\par\nobreak\vspace{-5pt}
\begingroup
\allowdisplaybreaks
\begin{small}
	\begin{equation} \label{eq:Paux}
		F_{\mathrm{b}}(s) = F_{\mathrm{dc}}(s) +P_{\mathrm{aux}}\cdot \frac{\mathrm{d}t}{\mathrm{d}s}(s),
	\end{equation}
\end{small}%
\endgroup 
where $F_{\mathrm{b}}(s)$ is the force equivalent of the battery power at the terminals.

The battery efficiency is mostly determined by its internal resistance $R_{\mathrm{0}}$ and open-circuit voltage $V_{\mathrm{oc}}$. We derive the battery losses from a Thévenin model~\cite{GuzzellaSciarretta2007} as
\par\nobreak\vspace{-5pt}
\begingroup
\allowdisplaybreaks
\begin{small}
	\begin{equation}
		P_{\mathrm{i}}(s) = \frac{1}{P_{\mathrm{sc}}} \cdot P_{\mathrm{i}}^2(s)+P_{\mathrm{b}}(s),
		\label{eq:Pi}
	\end{equation}
\end{small}%
\endgroup 
where $P_{\mathrm{sc}} = \frac{V_{\mathrm{oc}}^2}{R_{\mathrm{0}}}$ is the short-circuit power~\cite{BorsboomFahdzyanaEtAl2020}, which can be obtained from manufacturer data and which we assume to be constant. $P_{\mathrm{b}}(s)$ is the power at the battery terminals and $P_{\mathrm{i}}(s)$ is the internal battery power, which ultimately dictates a change in battery energy. Translating \eqref{eq:Pi} to forces and relaxing results in
\par\nobreak\vspace{-5pt}
\begingroup
\allowdisplaybreaks
\begin{small}
	\begin{equation}\label{eq:eff_bat}
		(F_{\mathrm{i}}(s)-F_{\mathrm{b}}(s))\cdot \frac{\mathrm{d}t}{\mathrm{d}s}(s)\cdot P_{\mathrm{sc}} \geq F_{\mathrm{i}}^2(s),
	\end{equation}
\end{small}%
\endgroup
where $F_{\mathrm{i}}(s)$ is the internal battery force and $F_{\mathrm{b}}(s)$ is the battery force at the terminals.

The energy consumption of the battery is modeled as
\par\nobreak\vspace{-5pt}
\begingroup
\allowdisplaybreaks
\begin{small}
	\begin{equation} \label{eq:dEbat_ds}
		\frac{\mathrm{d}}{\mathrm{d}s}E_{\mathrm{b}}(s) = -F_{\mathrm{i}}(s).
	\end{equation}
\end{small}%
\endgroup
We constrain the battery energy as
\par\nobreak\vspace{-5pt}
\begingroup
\allowdisplaybreaks
\begin{small}
	\begin{align}
		&E_{\mathrm{b}}(0) = E_\mathrm{b,0}, \\ 	
		E_{\mathrm{b,min}} \leq &E_{\mathrm{b}}(s) \leq E_{\mathrm{b,max}},
		\label{eq:Ebat_bound}
	\end{align} 
\end{small}%
\endgroup
where $E_\mathrm{b,0}$ is the initial battery energy. Furthermore, $E_{\mathrm{b,min}}$ and $E_{\mathrm{b,max}}$ correspond to the battery energy at the lower and upper \gls{acr:soe} bound, respectively. We leverage a lookup table with input charge time $t_\mathrm{charge}$ and output $E_\mathrm{b,0}$ for a given charging profile during pre-processing. 



\subsection{Low-level Optimization Problem}
\label{sec:Methodology - low-level optimization}
This section presents the minimum-stint-time control problem of the electric race car. Given a predefined stint length and charge time we formulate the control problem using the state variables $x = (E_{\mathrm{kin}},E_{\mathrm{b}})$ and the control variables $u = (F_{\mathrm{m}},F_{\mathrm{brake}})$ as follows:\\

	\begin{prob}[Minimum-stint-time Control Strategy]\label{prob:low-level}
		The minimum-stint-time control strategies are the solution of
		\par\nobreak\vspace{-5pt}
		\begingroup
		\allowdisplaybreaks
		\begin{small}
		\begin{equation*}
			\begin{aligned}
				&\min \int_{0}^{S_{\mathrm{stint}}} {\dtds(s)}\,\mathrm{d}s ,\\
				&\textnormal{s.t. }  \eqref{eq:lethargy}, \eqref{eq:F_rl}-\eqref{eq:P_em,max}, \eqref{eq:eff_EM}, \eqref{eq:eff_inverter}, \\
				& \qquad \eqref{eq:Paux}, \eqref{eq:eff_bat}- \eqref{eq:Ebat_bound}.\\
			\end{aligned}
		\end{equation*}
	\end{small}%
\endgroup
	\end{prob}
\noindent Since the constraint set and the cost function are convex, the low-level control problem is fully convex and therefore we can compute the globally optimal solution with standard nonlinear programming methods.

\section{High-level Race Optimization} \label{sec:high level}
In this section, we present the high-level maximum-race-distance control problem. First, we formulate the maximum-race-distance control problem that optimizes the stint length and charge time for a pre-defined number of pit stops. Second, we model the minimum stint time by leveraging the low-level control problem and optimizing for various combinations of stint length and initial battery energy.
Finally, we extend the maximum-race-distance control problem to allow joint optimization of the stint length, charge time, and number of pit stops. 
%
\subsection{Mixed-integer Control Problem}


We define the high-level control problem for a pre-defined number of pit stops in \emph{stint domain}, so that we have a fixed and finite optimization horizon. Here, each index in the optimization variables represents a stint. The goal is then to maximize the driven distance as the sum of all completed laps during the stints as
\par\nobreak\vspace{-5pt}
\begingroup
\allowdisplaybreaks
\begin{small}
	\begin{equation}
		\max S_{\mathrm{race}} = \max \sum_{k=0}^{N_{\mathrm{stops}}} S_{\mathrm{lap}}\cdot N_{\mathrm{laps}}(k),
	\end{equation}
\end{small}%
\endgroup
where $S_{\mathrm{race}}$ is the total race distance, $N_{\mathrm{stops}}$ is the pre-defined number of pit stops, $N_{\mathrm{laps}}(k)\in\sN, \  \forall~k \in \left[0,...,~N_{\mathrm{stops}}~-~1\right]$ is the stint length and $\sN$ the set of natural numbers, and $S_{\mathrm{lap}}$ is the length of one lap. Since the vehicle starts and stops at the pit box, the stint length should be an integer number of laps. As it is unlikely that the vehicle is exactly at the finish line when the race time limit is reached, we allow the final stint length to be a non-integer number of laps. This way, we have $N_{\mathrm{stops}}+1$ stints for $N_\mathrm{stops}$ pit stops and thus we have $N_\mathrm{stops}$ integer stint lengths and one final non-integer stint length.

The race can be divided into the car driving a stint and recharging the battery during pit-stops. Given the total race time $t_{\mathrm{race}}$, we can link it to the time to complete the stint $t_{\mathrm{stint}}(k) \geq 0$ and the time spent charging $t_{\mathrm{charge}}(k)\geq 0$ as
\par\nobreak\vspace{-5pt}
\begingroup
\allowdisplaybreaks
\begin{small}
	\begin{equation} \label{eq:t_race}
		t_{\mathrm{race}} = \sum_{k=0}^{N_{\mathrm{stops}}} t_{\mathrm{stint}}(k) + \sum_{k=1}^{N_{\mathrm{stops}}} t_{\mathrm{charge}}(k).
	\end{equation}
\end{small}%
\endgroup
We then decompose the total race into blocks consisting of a pit stop followed by a stint. By assuming that a stint is always energy limited, the charge time uniquely defines the initial battery energy for the subsequent stint and is not influenced by prior stints. To ensure that the battery is not overcharged, we apply an upper bound on the charge time through
\par\nobreak\vspace{-5pt}
\begingroup
\allowdisplaybreaks
\begin{small}
	\begin{equation}\label{eq:t_charge_max}
		t_\mathrm{charge}(k) \leq t_{\mathrm{charge,max}},
	\end{equation}
\end{small}%
\endgroup
where $t_{\mathrm{charge,max}}$ is the maximum charge time corresponding to a full battery, assuming that the battery is always charged starting from the lower energy bound. 
Since we start the race with a full battery, we set \mbox{$t_{\mathrm{charge}}(0)=t_{\mathrm{charge,max}}$} and do not count it in the objective.
Finally, the time to complete the stint is obtained by solving the low-level control problem, which we explain in the subsequent section.

\subsection{Stint Time Model}\label{sec:stinttime}
\begin{figure}
	\vspace{2mm}
	\centering
	\includegraphics[width=0.95\linewidth]{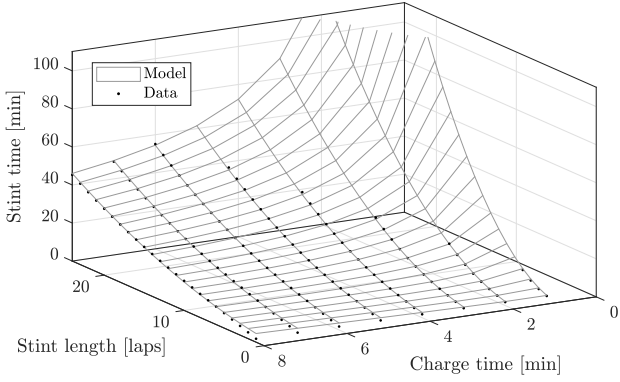}
	\caption{Fit of optimization data for a combination of stint lengths and charge times. The normalized RMSE of the fit is 0.80\% w.r.t. the maximum stint time.}
	\label{fig:t_stint_lut}
\end{figure}

In this section, we derive a method for modeling the stint time as a function of the stint length and charge time during the pit stop prior to the stint.
We solve the low-level control Problem~\ref{prob:low-level} for a combination of stint lengths and initial battery energy to obtain the respective achievable minimum stint time.
This way, we can create the lookup table with stint time as a function of stint length and charge time, as shown in Fig.~\ref{fig:t_stint_lut}. Thereby the charge time and initial battery energy are linked through a pre-defined charging current profile, cf. Section~\ref{sec:batt}.
As the stint time increases for larger stint lengths and shorter charge times, similar to the \gls{acr:em} loss fit in Section~\ref{sec:EM} above, we approximate the low-level optimization results via the continuous function
\par\nobreak\vspace{-5pt}
\begingroup
\allowdisplaybreaks
\begin{small}
	\begin{equation}
		t_{\mathrm{stint}}(k) = x_{\mathrm{s}}^{\top}(k) Q_{\mathrm{s}}x_{\mathrm{s}}(k),
		\label{eq:t_stint}
	\end{equation}
\end{small}%
\endgroup
where $x_{\mathrm{s}}(k) = \left[\frac{1}{\sqrt{t_{\mathrm{charge}}(k)}} \  \sqrt{t_{\mathrm{charge}}(k)} \  \frac{N_{\mathrm{laps}}(k)}{\sqrt{t_{\mathrm{charge}}(k)}}\right]^\top$ and $Q_{\mathrm{s}}$ is a symmetric positive semi-definite matrix of coefficients. The result of the fit is shown in Fig.~\ref{fig:t_stint_lut}. For a convex implementation, we relax and rewrite \eqref{eq:t_stint} to
\par\nobreak\vspace{-5pt}
\begingroup
\allowdisplaybreaks
\begin{small}
	\begin{equation}
		t_{\mathrm{stint}}(k)\cdot t_{\mathrm{charge}}(k) \geq y_{\mathrm{s}}(k)^\top Q_{\mathrm{s}}y_{\mathrm{s}}(k),
		\label{eq:t_stint2}
	\end{equation}
\end{small}%
\endgroup
where $y_{\mathrm{s}}(k) = \left[1 \ t_{\mathrm{charge}}(k) \  N_{\mathrm{laps}}(k)\right]^\top$, and convert this relaxation to a conic constraint~\cite{BoydVandenberghe2004} as
\par\nobreak\vspace{-5pt}
\begingroup
\allowdisplaybreaks
\begin{small}
	\begin{equation}
		t_{\mathrm{stint}}(k) + t_{\mathrm{charge}}(k) \geq 
		\begin{Vmatrix}
			2\cdot z_{\mathrm{s}}(k) \\
			t_{\mathrm{stint}}(k) - t_{\mathrm{charge}}(k)\\	
		\end{Vmatrix}_2,
		\label{eq:t_stint_cone}
	\end{equation}
\end{small}%
\endgroup
where $z_{\mathrm{s}}=C_{\mathrm{s}}y_{\mathrm{s}}(k)$ with $C_{\mathrm{s}}$ being the Cholesky factorization of $Q_{\mathrm{s}}$~\cite{BoydVandenberghe2004}. Since it is optimal to minimize stint time, this constraint will hold with equality at the optimum.



\par
\subsection{Optimal Pit Stop Strategy}
In the previous sections, we introduced the objective and constraints for the high-level control problem when optimizing the race strategy for a pre-defined number of pit stops. In this section, we apply some modifications in order to jointly optimize the stint lengths, charge times and number of pit stops, thereby removing the need to search over a large space of pre-defined number of pit stops.

We define a binary variable $b_{\mathrm{pit}}(k)$ that indicates whether pit stop and stint $k$ is taken or skipped as
\par\nobreak\vspace{-5pt}
\begingroup
\allowdisplaybreaks
\begin{small}
	\begin{equation}
		b_{\mathrm{pit}}(k) =
		\begin{cases}
			0, & \text{if stop and stint skipped} \\
			1, & \text{if stop and stint taken}, \\
		\end{cases}
	\end{equation}
\end{small}%
\endgroup
and include it in \eqref{eq:t_stint} via the big-M formulation~\cite{RichardsHow2005}
\par\nobreak\vspace{-5pt}
\begingroup
\allowdisplaybreaks
\begin{small}
	\begin{equation}
		t_{\mathrm{stint}}(k) \geq x_{\mathrm{s}}(k)^\top Q_{\mathrm{s}}x_{\mathrm{s}}(k) - M\cdot (1-b_{\mathrm{pit}}(k)),
		\label{eq:t_stint_M}
	\end{equation}
\end{small}%
\endgroup
where $M \gg t_{\mathrm{stint,max}}$. This way, we obtain the original constraint if $b_{\mathrm{pit}}(k)=1$ and we obtain a negative lower bound when $b_{\mathrm{pit}}(k)=0$ which, together with $t_{\mathrm{stint}}(k) \geq 0$, will push the $k$-th stint time to zero, hence skipping the stint.
We convert \eqref{eq:t_stint_M} to a cone as
\par\nobreak\vspace{-5pt}
\begingroup
\allowdisplaybreaks
\begin{small}
	\begin{equation} \label{eq:t_stint_cone_M}
		\begin{split}
			M\cdot (1-b_{\mathrm{pit}}(k)) + t_{\mathrm{stint}}(k) + t_{\mathrm{charge}}(k)  \geq  \\
			\begin{Vmatrix}
				2\cdot z_{\mathrm{s}}(k) \\
				M\cdot (1-b_{\mathrm{pit}}(k)) +t_{\mathrm{stint}}(k) - t_{\mathrm{charge}}(k)   \\	
			\end{Vmatrix}_2.\\
		\end{split}
	\end{equation}
\end{small}%
\endgroup
Hence, whenever a stint is skipped, the corresponding stint time and charge time will be zero if an optimal solution is obtained. To prevent the stint length from diverging to infinity whenever the stint is actually skipped, i.e., $b_\mathrm{pit}(k) = 0$, we define an upper bound on stint length as
\par\nobreak\vspace{-5pt}
\begingroup
\allowdisplaybreaks
\begin{small}
	\begin{equation} \label{eq:Nlaps_M}
		N_{\mathrm{laps}}(k) \leq N_{\mathrm{laps,max}} \cdot b_{\mathrm{pit}} (k),
	\end{equation}
\end{small}%
\endgroup
where $N_\mathrm{laps,max}$ is the maximum stint length that was used to obtain the lookup table. This will ensure $N_{\mathrm{laps}}(k)=0$ whenever $b_{\mathrm{pit}}(k)=0$.
Finally, we enforce $b_{\mathrm{pit}}(0)=1$ since the first stint at the start of the race is always taken, and place driven stints first as
\par\nobreak\vspace{-5pt}
\begingroup
\allowdisplaybreaks
\begin{small}
	\begin{equation} \label{eq:b_pit}
		b_{\mathrm{pit}}(k+1) \geq b_{\mathrm{pit}}(k), \quad  \forall k \in [1,N_{\mathrm{stops}}]
	\end{equation}
\end{small}%
\endgroup

\subsection{High-level Optimization Problem}
\label{sec:Methodology - high-level optimization}
This section presents the maximum-race-distance control problem of the electric race car. Given a predefined race time we formulate the control problem using the control variables $(t_{\mathrm{charge}},N_{\mathrm{laps}},b_{\mathrm{pit}})$ as follows:\\

	\begin{prob}[Maximum-race-distance Strategies]\label{prob:high-level}
		The maximum-race-distance strategies are the solution of
		\par\nobreak\vspace{-5pt}
		\begingroup
		\allowdisplaybreaks
		\begin{small}
		\begin{equation*}
			\begin{aligned}
				& \max \sum_{k=0}^{N_{\mathrm{stops}}} S_{\mathrm{lap}}\cdot N_{\mathrm{laps}}(k) ,\\
				&\textnormal{s.t. }  \eqref{eq:t_race}, \eqref{eq:t_charge_max}, \eqref{eq:t_stint_cone_M}-\eqref{eq:b_pit}. \\
			\end{aligned}
		\end{equation*}
	\end{small}%
\endgroup
	\end{prob}

\noindent Since Problem 2 can be solved with mixed-integer second-order conic programming solvers, we can guarantee global optimality upon convergence.

\section{Discussion}
A few comments are in order.
First, we assume that endurance racing tires do not degrade significantly and can be changed every stint due to the long pit stop time. Yet, the high-level control problem can be readily extended to capture these dynamics if the lookup table is devised accounting for tire degradation. 
Second, we assume that the time gained from starting the race from the grid compared to the pit lane is negligible on a full endurance race. Thus we do not separately optimize the first stint. Similarly, we do not separately optimize the final stint, since we assume that the vehicle can push the SoE below the lower limit to complete the final lap of the race, as battery degradation would no longer be an issue. 
Third, we assume that the cooling system can cope with the requested power from the battery and \gls{acr:em} and devote temperature modeling to future research. Yet again, the high-level control problem can capture temperature effects if the map is devised accounting for the temperatures and by assuming that the temperatures at the start of the stint are always the same.
Finally, it might occur that the vehicle can recuperate more energy, compared to what is needed to drive to the pit box, during braking before pit entry. Yet this amount of energy can be neglected, since it does not affect the \gls{acr:PO} and stint time.
%

%% file: Sections/Results.tex


\section{Results } \label{Results}
%
This section presents numerical results for both the low- and high-level control problem. We base our use case on the electric endurance race car of InMotion~\cite{InMotion}, shown in Fig.~\ref{fig:inmotion-car}, performing a 10 lap stint at the Zandvoort circuit for the low-level control problem and a \unit[6]{h} race at the same circuit for the high-level control problem. First, we discuss the numerical solutions for both control problems. Second, we validate the high-level control problem by comparing the optimal race strategy against fixed-pit-stop-number strategies and calculate the theoretical optimal combinations of stint length and charge time.
\par 
For the discretization of the model, we apply the Euler Forward method except for the lethargy, where we apply the trapezoidal method, with a fixed step-size of $\Delta s = \unit[4]{m}$. We parse the low-level control problem with CasADi~\cite{AnderssonGillisEtAl2019} and solve it using IPOPT~\cite{WachterBiegler2006} combined with the MA57 linear solver~\cite{HSL}, whilst we parse the high-level control problem with YALMIP~\cite{Loefberg2004} and solve it using MOSEK~\cite{ApS2017}.
We perform the numerical optimization on an Intel Core i7-4710MQ \unit[2.5]{GHz} processor and \unit[8]{GB} of RAM. Thereby, the computation time for solving the low-level problem was about \unit[4.6]{s} of parsing and \unit[25]{s} of solving, whereas the high-level problem needed \unit[0.04]{s} of parsing and \unit[0.57]{s} of solving.

\subsection{Low-level Optimization}
In this section, we compute the optimal trajectories for a stint of 10 laps around the Zandvoort circuit. We set the initial battery capacity to the upper bound corresponding to a \unit[7.5]{min} charge time. The total stint time is about \unit[946]{s} with an average flying lap time of \unit[93]{s}.   
\par
\begin{figure}[!tb]
	\centering
	\input{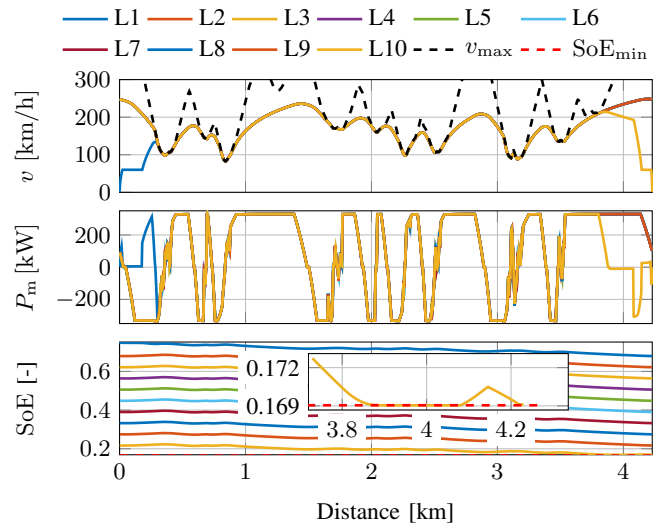}
	\caption{Velocity, EM power and battery SoE trajectories per lap for a 10 lap stint. The battery energy is an active constraint, thus the stint is energy limited. The \gls{acr:em} power shows a gradual decrease at high velocities, thus indicating energy management. The zoom window corresponds to the final \unit[500]{m} of the stint.}
	\label{fig:lap_results}
\end{figure}

 
The velocity profile together with the EM power  and \gls{acr:soe} per lap is shown in Fig.~\ref{fig:lap_results}.
First, we observe that the velocity profiles of consecutive free-flow laps are equivalent, as there are no lap-dependent dynamics.
Second, the \gls{acr:em} power decreases gradually before the vehicle reaches a corner and full regenerative braking is applied, which defines the optimal \gls{acr:PO}. Finally, we observe that the pit lane speed limit is adhered to, but the power at pit exit and pit entry are slightly different. From the lower plot, we notice that the lower battery limit is reached before the end of the stint, indicating that the recuperated energy during pit entry is larger than the required energy for driving through the pit lane.
However, this does not affect the stint time or the \gls{acr:PO}.

\subsection{High-level Optimization}
This section presents the optimal race strategy in terms of number of pit stops, stint length and charge time, and we compare it against the strategies optimized for a fixed number of pit stops. 
We select a 6 hour race, yet longer races can be solved as well with our approach, considering the very low computational times needed by our high-level framework to converge.
To link the initial battery energy $E_\mathrm{b,0}$ to the charge time $t_\mathrm{charge}$, we apply constant current charging starting from the lower \gls{acr:soe} limit.
%
\par
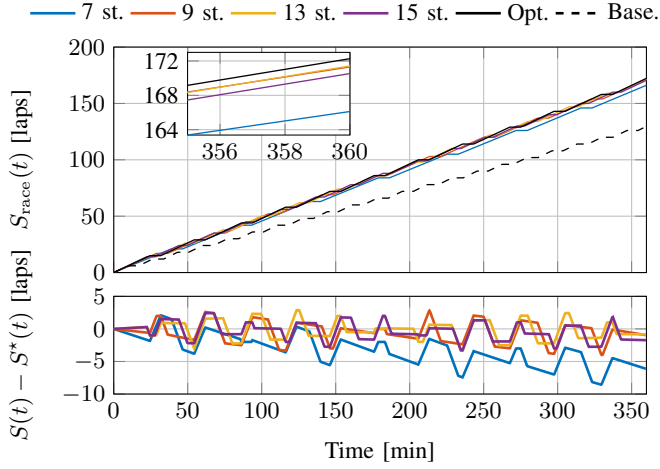
\begin{figure}[!tb]
	\centering
	\input{./Figures/cum_strategy.tex}
	\caption{Evolution of the completed laps as a function of time for the optimal strategy (black) and the strategies optimized for a fixed number of pit stops. The zoom window corresponds to the final \unit[5]{min} of the race and illustrates the difference in race distance between the optimized strategies. Jointly optimizing the number of stints can significantly outperform other strategies by multiple laps.}
	\label{fig:cumulative_strategy}
\end{figure}

 
Fig.~\ref{fig:cumulative_strategy} shows the evolution of the completed laps as a function of time for various fixed-pit-stop-number strategies. We observe that the optimal strategy of \unit[11]{stops} results in the largest number of completed laps, thereby confirming that it is indeed optimal in terms of number of pit stops. The difference in covered race length between the optimal and fixed-pit-stop-number strategies can exceed multiple laps and hence significantly affect the final race outcome in terms of finishing position, highlighting the importance of jointly optimizing the number of pit stops. Lastly, to show the importance of the bi-level approach, we compare the optimal strategy against a baseline \emph{flat-out} strategy whereby no energy management is applied to limit energy consumption, but the EMs are rather operated at maximum power whenever possible.
This results in \unit[6]{lap} stints and a total race distance of \unit[132]{laps}, whilst the globally optimal solution is about \unit[172]{laps}, which is about 30\% better compared to the baseline. 

\begin{figure}[!tb]
	\centering
	\input{./Figures/stint_strategy.tex}
	\caption{Optimal race strategy (black) in terms of stint length, charge time and stint time with $t_{\mathrm{charge,max}}=\unit[7.5]{min}$. For comparison, we show other optimal fixed pit stops strategies together with the relaxed solution in gray. Stint length, charge- and stint time are related and the optimal integer solution minimizes the differences to the relaxed solution.}
	\label{fig:stint_strategy}
\end{figure}
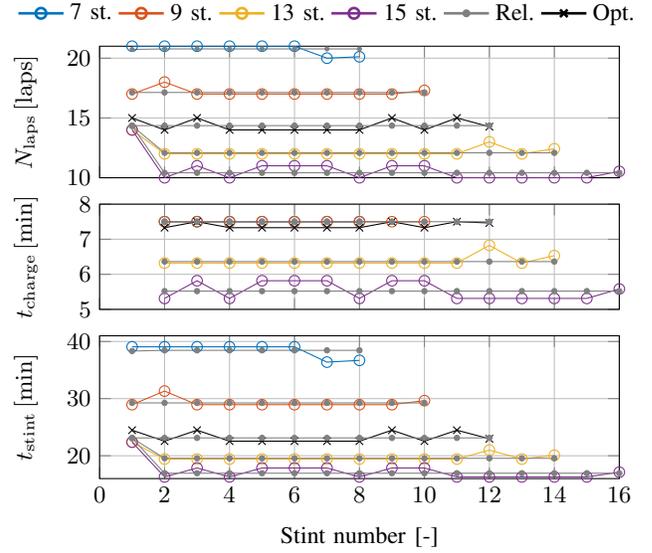

Fig.~\ref{fig:stint_strategy} shows the individual stints in terms of length and charge time, together with the relaxed non-integer solution.
We can conclude that a constant stint length over the race is optimal, since all stints in the relaxed solutions are equal, with the only exception being the first stints.
In this use case, the optimal integer solution consists of the stint lengths that minimize the difference to the relaxed solution, namely, of a stint length between 14 and 15 laps together with a charge time of almost \unit[7.5]{min} and 11 pit stops in total.
For strategies with more stops, both the stint lengths and charge times are reduced, thus showing that partly charging the battery is optimal for more than \unit[11]{stops}. For strategies with fewer than \unit[11]{stops}, the charge time is already maximized and no compensation is possible for increasing stint lengths.
From the aforementioned observations, we conclude that the stint length, stint time and charge time are closely related in the case of an optimal solution. Thereby, all the stints consist of a unique and lap-wise equal globally optimal \gls{acr:PO}.


\subsection{Validation}
%
In this section, we validate the numerical combinations of stint length and charge time for the various strategies. To this end, we calculate the average stint velocity $\overline{v}_{\mathrm{stint}}(k)$ for every strategy as
\par
\begin{figure}[!tb]
	\vspace{2mm}
	\centering
	\small
	\input{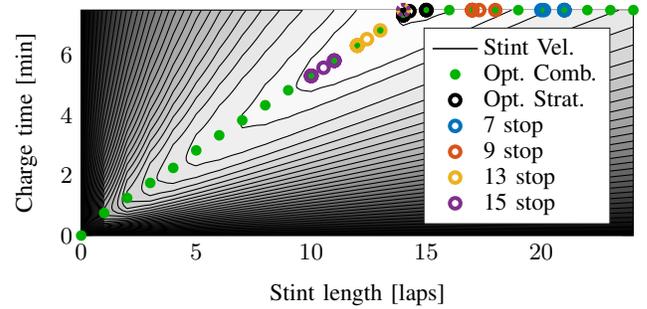}
	\caption{The average stint velocity for a combination of stint length and charge time together with the optimal combinations and actual numerical solutions. The optimal combinations of stint length and charge time show a clear (linear) correlation, to which the numerical solutions are aligned. The dashed circles indicate the first stints for the 13 and 15 stop strategies. }
	\label{fig:stint_velocity}
\end{figure}

\nobreak\vspace{-5pt}
\begingroup
\allowdisplaybreaks
\begin{small}
	\begin{equation}\label{eq:stint_velocity}
		\overline{v}_{\mathrm{stint}}(k) = \frac{S_{\mathrm{stint}}(k)}{t_{\mathrm{charge}}(k)+t_{\mathrm{stint}}(k)}, \quad \forall k > 0.
	\end{equation}
\end{small}%
\endgroup
Arguably, the globally optimal stint should maximize the average stint velocity. Fig.~\ref{fig:stint_velocity} shows the average stint velocity for all possible combinations of stint length and charge time together with the theoretical optimal charge times that maximize the average stint velocity for a given stint length, to which we refer as the optimal combinations. These optimal combinations show an almost linear relation between charge time and stint length until the maximum charge time is reached. 
The globally optimal stint consists of \unit[15]{laps} and \unit[7.5]{min} charging, which is the exact same combination that we obtained as the optimal strategy in the previous section. Furthermore, we observe that the average stint velocity decreases in sensitivity around the optimal combinations for increasing stint length and charge time, until the maximum charge time is reached. Thereby, increasing the stint length beyond \unit[15]{laps} quickly becomes less favorable, explaining why the \unit[7]{stop} strategy is significantly worse than the others. 
Finally, we note that the numerical solutions align well with the theoretically optimal combinations.
The outliers at \unit[14]{laps} and \unit[7.5]{min} charging correspond to the first stints, for which the charge time is not part of the race and thus the calculation of the stint velocity in~\eqref{eq:stint_velocity} is not valid.


%% file: Figures/cum_strategy.tex
%
%
\definecolor{mycolor1}{rgb}{0.00000,0.44700,0.74100}%
\definecolor{mycolor2}{rgb}{0.85000,0.32500,0.09800}%
\definecolor{mycolor3}{rgb}{0.92900,0.69400,0.12500}%
\definecolor{mycolor4}{rgb}{0.49400,0.18400,0.55600}%
\begin{tikzpicture}
\small
\begin{groupplot}[%
group style={
	group size=1 by 2,
	vertical sep=9pt},
width=0.82\columnwidth,
at={(0,0)},
scale only axis,
xmin=0,
xmax=360,
axis background/.style={fill=white},
xmajorgrids,
ymajorgrids,
xlabel={Distance [km]}
]
\nextgroupplot[%
height=3.0cm,
xticklabels={,,},
ylabel style={align=center},
ylabel={$S_{\mathrm{race}}(t)$ [laps]},
ymin=0,
ymax=200,
legend style={at={(-0.175,1.15)}, anchor=west, legend cell align=left, align=left, draw=none, fill=none},
legend entries={7 st., 9 st., 13 st., 15 st., Opt., Base.},
legend columns=6,
legend image post style={line width=1pt}
]
\addplot [color=mycolor1, line width=0.5pt]
table[row sep=crcr]{%
	0	0\\
	39.0691256637369	21\\
	46.5691256640932	21\\
	85.6382513306263	42\\
	93.1382513299534	42\\
	132.207376991125	63\\
	139.7073769914	63\\
	178.776502657552	84\\
	186.276502657367	84\\
	225.345628321083	105\\
	232.845628320694	105\\
	271.914753983344	126\\
	279.414753984069	126\\
	315.802798462681	145.999999954122\\
	323.302798462712	145.999999954122\\
	360.000000020774	166.117227728288\\
};

\addplot [color=mycolor2, line width=0.5pt]
table[row sep=crcr]{%
	0	0\\
	28.9408524852063	17\\
	36.440852485012	17\\
	67.7647605012081	35\\
	75.2647605005573	35\\
	104.205612992396	52\\
	111.705612991627	52\\
	140.64646548301	69\\
	148.146465482262	69\\
	177.087317973736	86\\
	184.58731797292	86\\
	213.528170464118	103\\
	221.028170463606	103\\
	249.969022954554	120\\
	257.469022953986	120\\
	286.409875448666	137\\
	293.909875448185	137\\
	322.850727940681	154\\
	330.350727940151	154\\
	360.000000000084	171.301664556813\\
};

\addplot [color=mycolor3, line width=0.5pt]
table[row sep=crcr]{%
	0	0\\
	22.3877370871477	14\\
	28.7084625806317	14\\
	48.1317879167995	26\\
	54.4525134102834	26\\
	73.8758387464512	38\\
	80.1965642399351	38\\
	99.6198895761029	50\\
	105.940615069587	50\\
	125.363940405755	62\\
	131.684665899642	62\\
	151.10799123705	74\\
	157.428716226919	74\\
	176.852040413614	85.9999992105276\\
	183.172765907098	85.9999992105276\\
	202.596091243265	97.9999992105276\\
	208.916816736749	97.9999992105276\\
	228.340142072917	109.999999210528\\
	234.660867566401	109.999999210528\\
	254.084192902569	121.999999210528\\
	260.404918396053	121.999999210528\\
	279.82824373222	133.999999210528\\
	286.653075302404	133.999999210528\\
	307.64124791094	146.999999210528\\
	313.961973404736	146.999999210528\\
	333.385298741864	158.999999210528\\
	339.918158034547	158.999999210528\\
	359.999952404505	171.420820339287\\
};

\addplot [color=mycolor4, line width=0.5pt]
table[row sep=crcr]{%
	0	0\\
	22.3877370859863	14\\
	27.7003439436972	14\\
	43.9941497874458	24\\
	49.8107980961174	24\\
	67.6693290263189	35\\
	72.9819358840299	35\\
	89.2757417277784	45\\
	95.09239003645	45\\
	112.950920966652	56\\
	118.767569275323	56\\
	136.626100205525	67\\
	142.442748514742	67\\
	160.301279446621	78\\
	165.613886304332	78\\
	181.907692148081	88\\
	187.724340456752	88\\
	205.582871386954	99\\
	211.399519695625	99\\
	229.258050625827	110\\
	234.570657483538	110\\
	250.864463327286	120\\
	256.177070184997	120\\
	272.470876028746	130\\
	277.783482886457	130\\
	294.077288730205	140\\
	299.389895587916	140\\
	315.683701431665	150\\
	320.996308525775	150\\
	337.290114578042	159.999999888385\\
	342.871542531436	159.999999888385\\
	359.999854891616	170.533403810012\\
};

\addplot [color=black, line width=0.5pt]
table[row sep=crcr]{%
	0	0\\
	24.472767043101	15.000000004069\\
	31.801724490555	15.000000004069\\
	54.3547887884052	28.9999996751076\\
	61.8547887884276	28.9999996751076\\
	86.3275558274477	43.999999675229\\
	93.6565134543895	43.999999675229\\
	116.209578253522	57.999999675229\\
	123.538535683302	57.999999675229\\
	146.091599935391	71.9999993149456\\
	153.42055736517	71.9999993149456\\
	175.973621617259	85.999998954662\\
	183.302579047038	85.999998954662\\
	205.855643299128	99.9999985943786\\
	213.184600728907	99.9999985943786\\
	235.737664980996	113.999998234095\\
	243.237664982127	113.999998234095\\
	267.710432027496	128.999998234217\\
	275.039389457275	128.999998234217\\
	297.592453709364	142.999997873933\\
	305.092453709458	142.999997873933\\
	329.565220750736	157.999997874055\\
	337.028858631157	157.999997874055\\
	359.999981806518	172.267143573212\\
};
\addplot [color=black, dashed, line width=0.5pt]
table[row sep=crcr]{%
	0	0\\
	9.37155833333333	6.0\\
	16.8715583333333	6\\
	26.2431166666667	12\\
	33.7431166666667	12\\
	43.114675	18\\
	50.614675	18\\
	59.9862333333334	24\\
	67.4862333333334	24\\
	76.8577916666667	30\\
	84.3577916666667	30\\
	93.72935	36\\
	101.22935	36\\
	110.600908333333	42\\
	118.100908333333	42\\
	127.472466666667	48\\
	134.972466666667	48\\
	144.344025	54\\
	151.844025	54\\
	161.215583333333	60\\
	168.715583333333	60\\
	178.087141666667	66\\
	185.587141666667	66\\
	194.9587	72\\
	202.4587	72\\
	211.830258333333	78\\
	219.330258333333	78\\
	228.701816666667	84\\
	236.201816666667	84\\
	245.573375	90\\
	253.073375	90\\
	262.444933333333	96\\
	269.944933333333	96\\
	279.316491666667	102\\
	286.816491666667	102\\
	296.18805	108\\
	303.68805	108\\
	313.059608333333	114\\
	320.559608333333	114\\
	329.931166666667	120\\
	337.431166666667	120\\
	346.802725	126\\
	354.302725	126\\
	363.674283333333	132\\
};
\coordinate (zoom) at (axis cs:50,122);

\nextgroupplot[%
height=1.3cm,
ymin=-10,
ymax=5,
ylabel style={align=center},
ylabel={$S(t)-S^\star(t)$ [laps]},
xlabel={Time [min]},
axis background/.style={fill=white},
legend style={legend cell align=left, align=left, draw=white!15!black}
]
\addplot [color=mycolor1, line width=1pt]
table[row sep=crcr]{%
	0	0\\
	24.1441441441441	-1.82088776571919\\
	24.5045045045045	-1.8286124735763\\
	31.7117117117117	2.04532503539212\\
	32.0720720720721	2.07120143188831\\
	38.9189189189189	1.50120571891574\\
	39.2792792792793	1.35824652067083\\
	46.4864864864865	-3.11568647563718\\
	46.8468468468469	-3.19010553763025\\
	54.0540540540541	-3.79010102524711\\
	54.4144144144144	-3.78308770315954\\
	61.6216216216216	0.0908498055316045\\
	61.981981981982	0.206586640718569\\
	85.4054054054054	-1.55994626583924\\
	85.7657657657658	-1.65566382992552\\
	86.1261261261261	-1.87653813469177\\
	86.4864864864865	-1.99999967522905\\
	92.972972972973	-1.99999967522905\\
	93.3333333333333	-1.89514137595444\\
	93.6936936936937	-1.724524439069\\
	116.036036036036	-3.58451046665317\\
	116.396396396396	-3.49854134644153\\
	123.243243243243	0.1816992873201\\
	123.603603603604	0.335004717494542\\
	131.891891891892	-0.35499009158201\\
	132.252252252252	-0.409110714967312\\
	139.459459459459	-4.8830437103436\\
	139.81981981982	-5.04630134704132\\
	145.945945945946	-5.55629751069148\\
	146.306306306306	-5.45301654401834\\
	153.153153153153	-1.77277591072578\\
	153.513513513513	-1.63678232690296\\
	175.855855855856	-3.49676833550944\\
	176.216216216216	-3.37617552250941\\
	178.738738738739	-2.02029739445425\\
	179.099099099099	-1.99999895466203\\
	183.063063063063	-1.99999895466203\\
	183.423423423423	-2.0750140926026\\
	185.945945945946	-3.64089064098425\\
	186.306306306306	-3.84856756630074\\
	205.765765765766	-5.46855537959624\\
	206.126126126126	-5.33065072025727\\
	212.972972972973	-1.65041008673523\\
	213.333333333333	-1.54904019555414\\
	225.225225225225	-2.53903274812359\\
	225.585585585586	-2.69801167078566\\
	232.792792792793	-7.17194466616189\\
	233.153153153153	-7.23034400322882\\
	235.675675675676	-7.44034242343383\\
	236.036036036036	-7.28512591776268\\
	243.243243243243	-3.41460747105032\\
	267.387387387387	-5.23549523114303\\
	267.747747747748	-5.23980087794934\\
	271.711711711712	-3.10913524795745\\
	272.072072072072	-2.9999982342166\\
	274.954954954955	-2.9999982342166\\
	275.315315315315	-3.1712814740448\\
	279.279279279279	-5.63194462150176\\
	279.63963963964	-5.73203712695516\\
	297.297297297297	-6.98797562943531\\
	297.657657657658	-6.9731311416669\\
	304.864864864865	-3.01182610648658\\
	305.225225225225	-2.89513999477987\\
	315.675675675676	-3.55660253139797\\
	316.036036036036	-3.70760620223894\\
	323.243243243243	-8.12509229715585\\
	323.603603603604	-8.18106668370564\\
	329.369369369369	-8.55429075257666\\
	329.72972972973	-8.47678539423282\\
	336.936936936937	-4.52582938540468\\
	337.297297297297	-4.49500628239633\\
	359.63963963964	-6.12365857081323\\
};

\addplot [color=mycolor2, line width=1pt]
table[row sep=crcr]{%
	0	0\\
	24.1441441441441	-0.616188642844406\\
	24.5045045045045	-0.605932766778039\\
	28.8288288288288	1.93419674568503\\
	29.1891891891892	1.99999999593103\\
	31.7117117117117	1.99999999593103\\
	32.0720720720721	1.83217951697884\\
	36.3963963963964	-0.852180280805953\\
	36.7567567567568	-0.89434540242047\\
	54.0540540540541	-1.69204938585926\\
	54.4144144144144	-1.67165512237921\\
	61.6216216216216	2.46990121416263\\
	61.981981981982	2.59901899074214\\
	67.7477477477478	2.37827518371608\\
	68.1081081081081	2.16717710361917\\
	74.954954954955	-2.029434686939\\
	75.3153153153153	-2.22061284328908\\
	86.1261261261261	-2.49651820657311\\
	86.4864864864865	-2.40830228778697\\
	93.3333333333333	1.61356943935795\\
	93.6936936936937	1.80216696010564\\
	104.144144144144	1.45361042759657\\
	104.504504504504	1.26602088320743\\
	111.711711711712	-3.20432970042043\\
	116.036036036036	-3.34855998969454\\
	116.396396396396	-3.24461028561734\\
	123.243243243243	0.777261441590895\\
	123.603603603604	0.948547455630944\\
	140.540540540541	0.38364550485386\\
	140.900900900901	0.222169690581154\\
	145.945945945946	-2.90958340618226\\
	146.306306306306	-2.99999931494557\\
	148.108108108108	-2.99999931494557\\
	148.468468468468	-2.81085316705241\\
	153.153153153153	-0.0590461958133233\\
	153.513513513513	0.0949279718990965\\
	175.855855855856	-0.650261835550168\\
	176.216216216216	-0.511688438660542\\
	176.936936936937	-0.0883335200083479\\
	177.297297297297	1.04533796729811e-06\\
	183.063063063063	1.04533796729811e-06\\
	183.423423423423	-0.0750140926026006\\
	184.504504504504	-0.746104041909007\\
	184.864864864865	-0.806768273593548\\
	205.765765765766	-1.50388131915503\\
	206.126126126126	-1.34799607593646\\
	212.972972972973	2.67387565129741\\
	213.333333333333	2.79322612635809\\
	213.693693693694	2.68397777158521\\
	220.900900900901	-1.78995522379103\\
	221.261261261261	-1.87673318530125\\
	235.675675675676	-2.35750080285663\\
	236.036036036036	-2.18430371330936\\
	243.243243243243	2.04582641092549\\
	249.72972972973	1.88028319410358\\
	250.09009009009	1.79997091407171\\
	257.297297297297	-2.61751518010675\\
	257.657657657658	-2.727584526219\\
	267.387387387387	-2.97589935218883\\
	267.747747747748	-2.9622244151463\\
	274.954954954955	1.2713247709064\\
	275.315315315315	1.31171899038083\\
	286.126126126126	0.951143276395328\\
	286.486486486487	0.894122383121953\\
	293.693693693694	-3.57981061225428\\
	294.054054054054	-3.71881603530773\\
	297.297297297297	-3.82698874935966\\
	297.657657657658	-3.79853205403168\\
	304.864864864865	0.43501713234042\\
	305.225225225225	0.56531545160675\\
	322.522522522522	0.123866871099096\\
	322.882882882883	0.0957820534682128\\
	329.369369369369	-3.87995543195706\\
	329.72972972973	-3.9999978740546\\
	330.09009009009	-3.9999978740546\\
	330.45045045045	-3.9418053687387\\
	336.936936936937	-0.15665296720141\\
	337.297297297297	-0.113091420104695\\
	359.63963963964	-0.951960175038266\\
};

\addplot [color=mycolor3, line width=1pt]
table[row sep=crcr]{%
	0	0\\
	22.3423423423424	0.277405847025818\\
	22.7027027027027	0.0849187983866386\\
	24.1441441441441	-0.79857842076342\\
	24.5045045045045	-1.00000000406897\\
	28.4684684684685	-1.00000000406897\\
	28.8288288288288	-0.925636065177628\\
	31.7117117117117	0.855448995997619\\
	32.0720720720721	0.910264149692296\\
	47.927927927928	0.863579394278474\\
	48.2882882882883	0.76583027528585\\
	54.0540540540541	-2.81331612176058\\
	54.4144144144144	-2.99999967510757\\
	54.7747747747748	-2.8009021309453\\
	61.6216216216216	1.42917488934592\\
	61.981981981982	1.57385048174518\\
	73.8738738738739	1.63197430180782\\
	75.6756756756757	0.528816703528605\\
	80	-2.12167495366606\\
	80.3603603603603	-2.24135374164592\\
	86.1261261261261	-2.21317249555494\\
	86.4864864864865	-2.11399840344535\\
	93.3333333333333	2.11607861684581\\
	93.6936936936937	2.31563431091701\\
	99.4594594594595	2.29865803167394\\
	99.8198198198198	2.17407733450216\\
	105.945945945946	-1.62547222778056\\
	116.036036036036	-1.65518071645585\\
	116.396396396396	-1.5402728390585\\
	123.243243243243	2.68980418123266\\
	123.603603603604	2.87204836859286\\
	125.045045045045	2.86780430010521\\
	125.405405405405	2.84112563064014\\
	131.531531531532	-0.961717415429689\\
	131.891891891892	-1.05738697077464\\
	145.945945945946	-1.09876663908335\\
	146.306306306306	-0.966546915213939\\
	150.990990990991	1.92771630901109\\
	151.351351351351	2.00000068505443\\
	153.153153153153	2.00000068505443\\
	153.513513513513	1.94229739344081\\
	157.117117117117	-0.294669104247305\\
	157.477477477477	-0.488240376785143\\
	175.855855855856	-0.542352325047489\\
	176.216216216216	-0.392820756308481\\
	176.576576576577	-0.17018512513306\\
	176.936936936937	2.55865586495929e-07\\
	183.063063063063	2.55865586495929e-07\\
	183.423423423423	0.0798448107867102\\
	202.522522522523	0.0236109033254479\\
	202.882882882883	-0.154633969590861\\
	205.765765765766	-1.94420716774141\\
	206.126126126126	-1.99999938385093\\
	208.648648648649	-1.99999938385093\\
	209.009009009009	-1.94304171836524\\
	212.972972972973	0.505950240750735\\
	213.333333333333	0.636258889130204\\
	228.108108108108	0.592757187131895\\
	228.468468468468	0.512414341591466\\
	234.594594594595	-3.29042870447836\\
	234.954954954955	-3.33243408349051\\
	235.675675675676	-3.33455611773428\\
	236.036036036036	-3.15040085487004\\
	243.243243243243	1.29889273570427\\
	254.054054054054	1.35173257384372\\
	254.414414414414	1.1494784680923\\
	260.18018018018	-2.38451040725045\\
	260.540540540541	-2.52159546857865\\
	267.387387387387	-2.48813023775705\\
	267.747747747748	-2.4634971273702\\
	274.954954954955	1.98921552556789\\
	275.315315315315	2.04056791838656\\
	279.63963963964	2.02783571292366\\
	280	1.92066128948829\\
	286.486486486487	-2.10587840635043\\
	286.846846846847	-2.20955364573979\\
	297.297297297297	-2.22378438280259\\
	297.657657657658	-2.18379921209555\\
	304.864864864865	2.2803194818581\\
	305.225225225225	2.4221462765035\\
	307.387387387387	2.43613605621454\\
	307.747747747748	2.37250206332141\\
	313.873873873874	-1.38236111735802\\
	314.234234234234	-1.43502890250602\\
	329.369369369369	-1.36105313112364\\
	329.72972972973	-1.25845994058528\\
	333.333333333333	0.967896385773713\\
	333.693693693694	1.00000133647302\\
	336.936936936937	1.00000133647302\\
	337.297297297297	0.833276639039923\\
	339.81981981982	-0.733438061129277\\
	340.18018018018	-0.795190731707407\\
	359.63963963964	-0.845375743295222\\
};

\addplot [color=mycolor4, line width=1pt]
table[row sep=crcr]{%
	0	0\\
	22.3423423423424	0.277405847750629\\
	22.7027027027027	0.0849187983866386\\
	24.1441441441441	-0.79857842076342\\
	24.5045045045045	-1.00000000406897\\
	27.3873873873874	-1.00000000406897\\
	27.7477477477477	-0.970906860020818\\
	31.7117117117117	1.4618973613974\\
	32.0720720720721	1.51524090257413\\
	43.963963963964	1.43166412292044\\
	44.3243243243243	1.22649342325525\\
	49.7297297297297	-2.12895632397573\\
	50.0900900900901	-2.1806224777942\\
	54.0540540540541	-2.19967361556945\\
	54.4144144144144	-2.16439244071699\\
	61.6216216216216	2.2749021232724\\
	61.981981981982	2.41890681122425\\
	67.3873873873874	2.435263162723\\
	67.7477477477478	2.38805140838537\\
	72.7927927927928	-0.704188858341695\\
	73.1531531531531	-0.819981961843325\\
	86.1261261261261	-0.809552208785874\\
	86.4864864864865	-0.711849729194284\\
	89.009009009009	0.836298411708242\\
	89.3693693693694	1.00000032477095\\
	93.3333333333333	1.00000032477095\\
	93.6936936936937	0.976920386195218\\
	94.7747747747748	0.305830435896439\\
	95.1351351351352	0.108462719983777\\
	112.792792792793	0.0235985468776221\\
	113.153153153153	-0.102698719182797\\
	116.036036036036	-1.89227191997952\\
	116.396396396396	-1.99999967522905\\
	118.558558558559	-1.99999967522905\\
	118.918918918919	-1.90677554128092\\
	123.243243243243	0.756801197112679\\
	123.603603603604	0.938374480025402\\
	136.576576576577	0.876025303528934\\
	136.936936936937	0.682832838038109\\
	142.342342342342	-2.67261690849409\\
	142.702702702703	-2.73619422490498\\
	145.945945945946	-2.75178151921671\\
	146.306306306306	-2.62023269980136\\
	153.153153153153	1.59709713559243\\
	153.513513513513	1.76135857215741\\
	160	1.73018398353395\\
	160.36036036036	1.69206104783336\\
	165.405405405405	-1.43969204893006\\
	165.765765765766	-1.5701756931615\\
	175.855855855856	-1.6410893230784\\
	176.216216216216	-1.49302936538589\\
	181.621621621622	1.82443093654797\\
	181.981981981982	2.00000104533797\\
	183.063063063063	2.00000104533797\\
	183.423423423423	1.9249859073974\\
	187.387387387387	-0.535677240059556\\
	187.747747747748	-0.744956119298536\\
	205.405405405405	-0.829820276196472\\
	205.765765765766	-0.94420637826903\\
	206.126126126126	-0.999998594378553\\
	211.171171171171	-0.999998594378553\\
	211.531531531532	-0.918685624087516\\
	212.972972972973	-0.0308267112896488\\
	213.333333333333	0.0988110326424021\\
	229.189189189189	0.0226064835911757\\
	229.54954954955	-0.158674818242616\\
	234.234234234234	-3.06673126523714\\
	234.594594594595	-3.2757369881063\\
	235.675675675676	-3.28333487702594\\
	236.036036036036	-3.10065122667964\\
	243.243243243243	1.31921011353506\\
	250.810810810811	1.32529413735517\\
	251.171171171171	1.13734799994495\\
	255.855855855856	-1.73401796127104\\
	256.216216216216	-1.9308671661388\\
	267.387387387387	-1.92188598811867\\
	267.747747747748	-1.89872449024983\\
	272.432432432432	0.976407771426238\\
	272.792792792793	1.00000176578345\\
	274.954954954955	1.00000176578345\\
	275.315315315315	0.828718525955196\\
	277.477477477478	-0.51346137265773\\
	277.837837837838	-0.703798749643852\\
	294.054054054054	-0.81776708343881\\
	294.414414414414	-1.0272039117919\\
	297.297297297297	-2.81677710994245\\
	297.657657657658	-2.99999787393307\\
	299.099099099099	-2.99999787393307\\
	299.459459459459	-2.95730442820894\\
	304.864864864865	0.360155873725034\\
	305.225225225225	0.499940753801638\\
	315.675675675676	0.508342499910952\\
	316.396396396396	0.0715195388934831\\
	320.720720720721	-2.57897211805675\\
	321.081081081081	-2.74781895046874\\
	329.369369369369	-2.74115561853148\\
	329.72972972973	-2.64003404579893\\
	336.936936936937	1.78324625080256\\
	337.297297297297	1.83327731689735\\
	342.702702702703	-1.52396846917947\\
	343.063063063063	-1.63000541018988\\
	359.63963963964	-1.73145636999936\\
};
\end{groupplot}
%
%
\begin{axis}[%
width=0.25\columnwidth,
height=1.1cm,
at={(zoom)},
scale only axis,
xmin=354.999854891616,
xmax=360,
ymin=163.376172438064,
ymax=173,
ytick={164,168,172},
yticklabels={164,168,172},
axis background/.style={fill=white},
xmajorgrids,
ymajorgrids,
]
\addplot [color=mycolor1, line width=0.5pt]
table[row sep=crcr]{%
	354.399854891616	163.047255350329\\
	360.000000020774	166.117227728288\\
};

\addplot [color=mycolor2, line width=0.5pt]
table[row sep=crcr]{%
	354.399854891616	168.033731639595\\
	360.000000000084	171.301664556813\\
};

\addplot [color=mycolor3, line width=0.5pt]
table[row sep=crcr]{%
	354.399854891616	167.957095523179\\
	359.999952404505	171.420820339287\\
};

\addplot [color=mycolor4, line width=0.5pt]
table[row sep=crcr]{%
	354.399854891616	167.089569956407\\
	359.999854891616	170.533403810012\\
};

\addplot [color=black, line width=0.5pt]
table[row sep=crcr]{%
	354.399854891616	168.7889581132\\
	359.999981806518	172.267143573212\\
};
\end{axis}
\end{tikzpicture}%

%% file: Figures/stint_strategy.tex
%
%
\definecolor{mycolor1}{rgb}{0.00000,0.44700,0.74100}%
\definecolor{mycolor2}{rgb}{0.85000,0.32500,0.09800}%
\definecolor{mycolor3}{rgb}{0.92900,0.69400,0.12500}%
\definecolor{mycolor4}{rgb}{0.49400,0.18400,0.55600}%
\begin{tikzpicture}
\small
\begin{groupplot}[%
group style={
	group size=1 by 3,
	vertical sep=10pt},
width=0.8\columnwidth,
at={(0,0)},
scale only axis,
xmin=0,
xmax=16,
y label style={yshift=-0.25cm},
axis background/.style={fill=white},
xmajorgrids,
ymajorgrids,
xlabel={Distance [km]}
]
\nextgroupplot[%
height=1.75cm,
xticklabels={,,},
ylabel style={align=center},
ylabel={$N_{\mathrm{laps}} \left[\mathrm{laps}\right]$},
ymin=10,
ymax=21,
legend style={at={(-0.175,1.25)}, anchor=west, legend cell align=left, align=left, draw=none, fill=none},
legend entries={7 st., , 9 st., , 13 st., , 15 st., Rel., Opt.},
legend columns=6,
legend image post style={line width=1pt}
]
\addplot [color=mycolor1, mark=o, mark options={solid, mycolor1}]
table[row sep=crcr]{%
	1	21\\
	2	21\\
	3	21\\
	4	21\\
	5	21\\
	6	21\\
	7	19.9999999541218\\
	8	20.1172277741666\\
};

\addplot [color=gray, mark size=0.75pt, mark=*, mark options={solid, gray}]
table[row sep=crcr]{%
	1	20.7280749541801\\
	2	20.7725796730265\\
	3	20.7724516731433\\
	4	20.7723367624571\\
	5	20.7722966203103\\
	6	20.7721405915019\\
	7	20.7719170626188\\
	8	20.7722926386172\\
};

\addplot [color=mycolor2, mark=o, mark options={solid, mycolor2}]
table[row sep=crcr]{%
	1	17\\
	2	18\\
	3	17\\
	4	17\\
	5	17\\
	6	17\\
	7	17\\
	8	17\\
	9	17\\
	10	17.3016645568125\\
};

\addplot [color=gray, mark size=1pt, mark=*, mark options={solid, gray}]
table[row sep=crcr]{%
	1	17.1339025126195\\
	2	17.1315874760143\\
	3	17.1317426037473\\
	4	17.1313944399184\\
	5	17.1315283273994\\
	6	17.1316821635061\\
	7	17.1312131745454\\
	8	17.1351447449902\\
	9	17.1300725677954\\
	10	17.1328780767445\\
};

\addplot [color=mycolor3, mark=o, mark options={solid, mycolor3}]
table[row sep=crcr]{%
	1	14\\
	2	12\\
	3	12\\
	4	12\\
	5	12\\
	6	12\\
	7	11.9999992105276\\
	8	12\\
	9	12\\
	10	12\\
	11	12\\
	12	13\\
	13	12\\
	14	12.420821128759\\
};
\addplot [color=gray, mark size=1pt, mark=*, mark options={solid, gray}]
table[row sep=crcr]{%
	1	14.3378097480806\\
	2	12.0835207490942\\
	3	12.0835207490942\\
	4	12.0835207490942\\
	5	12.0835207490942\\
	6	12.0835207490942\\
	7	12.0835207490942\\
	8	12.0835207490942\\
	9	12.0835207490942\\
	10	12.0835207490942\\
	11	12.0835207490942\\
	12	12.0835207490942\\
	13	12.0835207490942\\
	14	12.0835207491115\\
};

\addplot [color=mycolor4, mark=o, mark options={solid, mycolor4}]
table[row sep=crcr]{%
	1	14\\
	2	10\\
	3	11\\
	4	10\\
	5	11\\
	6	11\\
	7	11\\
	8	10\\
	9	11\\
	10	11\\
	11	10\\
	12	10\\
	13	10\\
	14	10\\
	15	9.99999988838506\\
	16	10.5334039216268\\
};

\addplot [color=gray, mark size=1pt, mark=*, mark options={solid, gray}]
table[row sep=crcr]{%
	1	14.3361745153988\\
	2	10.4133362086796\\
	3	10.4133362086796\\
	4	10.4133362086796\\
	5	10.4133362086796\\
	6	10.4133362086796\\
	7	10.4133362086796\\
	8	10.4133362086796\\
	9	10.4133362086796\\
	10	10.4133362086796\\
	11	10.4133362086796\\
	12	10.4133362086796\\
	13	10.4133362086796\\
	14	10.4133362086796\\
	15	10.4133362086796\\
	16	10.4133362086987\\
};

\addplot [color=black, mark=x, mark options={solid, black}]
table[row sep=crcr]{%
	1	15.000000004069\\
	2	13.9999996710386\\
	3	15.0000000001215\\
	4	14\\
	5	13.9999996397165\\
	6	13.9999996397165\\
	7	13.9999996397165\\
	8	13.9999996397165\\
	9	15.0000000001215\\
	10	13.9999996397165\\
	11	15.0000000001215\\
	12	14.2671456991575\\
};

\addplot [color=gray, mark size=1pt, mark=*, mark options={solid, gray}]
table[row sep=crcr]{%
	1	14.3590800525563\\
	2	14.3590800928814\\
	3	14.3590800983487\\
	4	14.3590800963931\\
	5	14.3590800928814\\
	6	14.3590800928814\\
	7	14.3590800928814\\
	8	14.3590800928814\\
	9	14.3590800928814\\
	10	14.3590800928814\\
	11	14.3590800855333\\
	12	14.3590800901152\\
};

\nextgroupplot[%
height=1.4cm,
ymin=5,
ymax=8,
ylabel style={align=center},
ylabel={$t_\mathrm{charge} \left[\mathrm{min}\right]$},
xlabel={Time [min]},
xticklabels={,,},
axis background/.style={fill=white},
legend style={legend cell align=left, align=left, draw=white!15!black}
]
\addplot [color=mycolor1, mark=o, mark options={solid, mycolor1}, x filter/.code={\pgfmathparse{#1+1}\pgfmathresult}]
table[row sep=crcr]{%
	1	7.50000000035629\\
	2	7.49999999932712\\
	3	7.50000000027496\\
	4	7.49999999981544\\
	5	7.49999999961104\\
	6	7.50000000072484\\
	7	7.50000000003119\\
};

\addplot [color=gray, mark size=1pt, mark=*, mark options={solid, gray}, x filter/.code={\pgfmathparse{#1+1}\pgfmathresult}]
table[row sep=crcr]{%
	1	7.49883884997275\\
	2	7.4988386980999\\
	3	7.49883865645301\\
	4	7.49883886897941\\
	5	7.49883834850518\\
	6	7.49883919832042\\
	7	7.4988389724624\\
};

\addplot [color=mycolor2, mark=o, mark options={solid, mycolor2}, x filter/.code={\pgfmathparse{#1+1}\pgfmathresult}]
table[row sep=crcr]{%
	1	7.4999999998058\\
	2	7.49999999934917\\
	3	7.49999999923175\\
	4	7.49999999925209\\
	5	7.49999999918397\\
	6	7.49999999948711\\
	7	7.49999999943193\\
	8	7.49999999951902\\
	9	7.49999999946997\\
};

\addplot [color=gray, mark size=1pt, mark=*, mark options={solid, gray}, x filter/.code={\pgfmathparse{#1+1}\pgfmathresult}]
table[row sep=crcr]{%
	1	7.49999337250781\\
	2	7.49999337605346\\
	3	7.49999338221015\\
	4	7.49999331687289\\
	5	7.49999350541323\\
	6	7.4999932681797\\
	7	7.49999394202178\\
	8	7.49999329064883\\
	9	7.49999361085982\\
};

\addplot [color=mycolor3, mark=o, mark options={solid, mycolor3}, x filter/.code={\pgfmathparse{#1+1}\pgfmathresult}]
table[row sep=crcr]{%
	1	6.32072549348398\\
	2	6.32072549348398\\
	3	6.32072549348398\\
	4	6.32072549348398\\
	5	6.32072549388745\\
	6	6.32072498986968\\
	7	6.32072549348398\\
	8	6.32072549348398\\
	9	6.32072549348398\\
	10	6.32072549348398\\
	11	6.82483157018319\\
	12	6.32072549379629\\
	13	6.53285929268359\\
};

\addplot [color=gray, mark size=1pt, mark=*, mark options={solid, gray}, x filter/.code={\pgfmathparse{#1+1}\pgfmathresult}]
table[row sep=crcr]{%
	1	6.36283963852752\\
	2	6.36283963852752\\
	3	6.36283963852752\\
	4	6.36283963852752\\
	5	6.36283963852752\\
	6	6.36283963852752\\
	7	6.36283963852752\\
	8	6.36283963852752\\
	9	6.36283963852752\\
	10	6.36283963852752\\
	11	6.36283963852752\\
	12	6.36283963852752\\
	13	6.36283963792327\\
};

\addplot [color=mycolor4, mark=o, mark options={solid, mycolor4}, x filter/.code={\pgfmathparse{#1+1}\pgfmathresult}]
table[row sep=crcr]{%
	1	5.31260685771099\\
	2	5.81664830867157\\
	3	5.31260685771099\\
	4	5.81664830867157\\
	5	5.81664830867157\\
	6	5.81664830921759\\
	7	5.31260685771099\\
	8	5.81664830867157\\
	9	5.81664830867157\\
	10	5.31260685771099\\
	11	5.31260685771099\\
	12	5.31260685771099\\
	13	5.31260685771099\\
	14	5.31260709411018\\
	15	5.58142795339377\\
};

\addplot [color=gray, mark size=1pt, mark=*, mark options={solid, gray}, x filter/.code={\pgfmathparse{#1+1}\pgfmathresult}]
table[row sep=crcr]{%
	1	5.52094508279949\\
	2	5.52094508279949\\
	3	5.52094508279949\\
	4	5.52094508279949\\
	5	5.52094508279949\\
	6	5.52094508279949\\
	7	5.52094508279949\\
	8	5.52094508279949\\
	9	5.52094508279949\\
	10	5.52094508279949\\
	11	5.52094508279949\\
	12	5.52094508279949\\
	13	5.52094508279949\\
	14	5.52094508279949\\
	15	5.52094508283191\\
};

\addplot [color=black, mark=x, mark options={solid, black}, x filter/.code={\pgfmathparse{#1+1}\pgfmathresult}]
table[row sep=crcr]{%
	1	7.32895744745406\\
	2	7.50000000002234\\
	3	7.3289576269418\\
	4	7.32895742977923\\
	5	7.32895742977923\\
	6	7.32895742977923\\
	7	7.32895742977923\\
	8	7.50000000113108\\
	9	7.32895742977923\\
	10	7.50000000009408\\
	11	7.46363788042128\\
};

\addplot [color=gray, mark size=1pt, mark=*, mark options={solid, gray}, x filter/.code={\pgfmathparse{#1+1}\pgfmathresult}]
table[row sep=crcr]{%
	1	7.49999999664277\\
	2	7.49999999059122\\
	3	7.49999999717329\\
	4	7.49999999664277\\
	5	7.49999999664277\\
	6	7.49999999664277\\
	7	7.49999999664277\\
	8	7.49999999664277\\
	9	7.49999999664277\\
	10	7.4999999921611\\
	11	7.49999999438598\\
};

\nextgroupplot[%
height=1.9cm,
ymin=16,
ymax=41,
ylabel style={align=center},
ylabel={$t_\mathrm{stint} \left[\mathrm{min}\right]$},
xlabel={Stint number [-]},
axis background/.style={fill=white},
legend style={legend cell align=left, align=left, draw=white!15!black}
]
\addplot [color=mycolor1, mark=o, mark options={solid, mycolor1}]
table[row sep=crcr]{%
	1	39.0691256637369\\
	2	39.0691256665331\\
	3	39.0691256611718\\
	4	39.0691256661513\\
	5	39.0691256637157\\
	6	39.0691256626508\\
	7	36.388044478612\\
	8	36.6972015580619\\
};

\addplot [color=gray, mark size=1pt, mark=*, mark options={solid, gray}, forget plot]
table[row sep=crcr]{%
	1	38.330616928892\\
	2	38.4546868665085\\
	3	38.4543413402711\\
	4	38.4540306462871\\
	5	38.4539227841116\\
	6	38.4535002477072\\
	7	38.452894375134\\
	8	38.4539111447876\\
};
\addplot [color=mycolor2, mark=o, mark options={solid, mycolor2}]
table[row sep=crcr]{%
	1	28.9408524852063\\
	2	31.323908016196\\
	3	28.9408524918383\\
	4	28.9408524913828\\
	5	28.9408524914739\\
	6	28.9408524911984\\
	7	28.9408524909481\\
	8	28.9408524946802\\
	9	28.9408524924964\\
	10	29.6492720599329\\
};

\addplot [color=gray, mark size=1pt, mark=*, mark options={solid, gray}, forget plot]
table[row sep=crcr]{%
	1	29.2541926375766\\
	2	29.2487699210184\\
	3	29.2491338628331\\
	4	29.2483168805342\\
	5	29.2486311899732\\
	6	29.2489918327084\\
	7	29.2478917572207\\
	8	29.2571168048195\\
	9	29.2452150358417\\
	10	29.2517980062525\\
};
\addplot [color=mycolor3, mark=o, mark options={solid, mycolor3}]
table[row sep=crcr]{%
	1	22.3877370871477\\
	2	19.4233253361677\\
	3	19.4233253361677\\
	4	19.4233253361677\\
	5	19.4233253361677\\
	6	19.4233253374075\\
	7	19.4233241866943\\
	8	19.4233253361677\\
	9	19.4233253361677\\
	10	19.4233253361677\\
	11	19.4233253361677\\
	12	20.9881726085365\\
	13	19.4233253371274\\
	14	20.0817943699574\\
};

\addplot [color=gray, mark size=1pt, mark=*, mark options={solid, gray}, forget plot]
table[row sep=crcr]{%
	1	23.0809694295435\\
	2	19.5540088671593\\
	3	19.5540088671593\\
	4	19.5540088671593\\
	5	19.5540088671593\\
	6	19.5540088671593\\
	7	19.5540088671593\\
	8	19.5540088671593\\
	9	19.5540088671593\\
	10	19.5540088671593\\
	11	19.5540088671593\\
	12	19.5540088671593\\
	13	19.5540088671593\\
	14	19.5540088646896\\
};
\addplot [color=mycolor4, mark=o, mark options={solid, mycolor4}]
table[row sep=crcr]{%
	1	22.3877370859862\\
	2	16.2938058437485\\
	3	17.8585309302015\\
	4	16.2938058437485\\
	5	17.8585309302015\\
	6	17.8585309302015\\
	7	17.8585309318788\\
	8	16.2938058437485\\
	9	17.8585309302015\\
	10	17.8585309302015\\
	11	16.2938058437485\\
	12	16.2938058437485\\
	13	16.2938058437485\\
	14	16.2938058437485\\
	15	16.2938060522673\\
	16	17.1283123601797\\
};

\addplot [color=gray, mark size=1pt, mark=*, mark options={solid, gray}, forget plot]
table[row sep=crcr]{%
	1	23.0775863964461\\
	2	16.9405491574283\\
	3	16.9405491574283\\
	4	16.9405491574283\\
	5	16.9405491574283\\
	6	16.9405491574283\\
	7	16.9405491574283\\
	8	16.9405491574283\\
	9	16.9405491574283\\
	10	16.9405491574283\\
	11	16.9405491574283\\
	12	16.9405491574283\\
	13	16.9405491574283\\
	14	16.9405491574283\\
	15	16.9405491574283\\
	16	16.9405491578388\\
};
\addplot [color=black, mark=x, mark options={solid, black}]
table[row sep=crcr]{%
	1	24.472767043101\\
	2	22.5530642978502\\
	3	24.4727670390202\\
	4	22.5530647991328\\
	5	22.5530642520892\\
	6	22.5530642520892\\
	7	22.5530642520892\\
	8	22.5530642520892\\
	9	24.4727670453687\\
	10	22.5530642520892\\
	11	24.4727670412774\\
	12	22.9711231753611\\
};

\addplot [color=gray, mark size=1pt, mark=*, mark options={solid, gray}, forget plot]
table[row sep=crcr]{%
	1	23.1249984059008\\
	2	23.1249984954317\\
	3	23.1249985067547\\
	4	23.1249985027047\\
	5	23.1249984954317\\
	6	23.1249984954317\\
	7	23.1249984954317\\
	8	23.1249984954317\\
	9	23.1249984954317\\
	10	23.1249984954317\\
	11	23.1249984802135\\
	12	23.1249984897027\\
};
 \end{groupplot}
\end{tikzpicture}%

%% file: Sections/Conclusion.tex
\section{Conclusion} \label{Conclusion}

In this paper, we devised a bi-level optimization framework to efficiently solve the maximum-distance endurance race strategy problem for a fully electric race car.
In order to tackle the large problem size stemming from the length of an endurance race, we decomposed the problem into separate stints which we solved by extending a minimum-lap-time convex optimization framework that can rapidly deliver the globally optimal solution to account for multiple laps and include more accurate force-based models.
This way, we were able to compute the optimal number of pit stops, the charging time per stop and the individual stint lengths via mixed-integer second-order conic programming with global optimality guarantees.
Our bi-level framework could solve the problem of a \SI{6}{\hour} race around the Zandvoort circuit with low computation times below \SI{1}{\second} for the high-level framework.
Our results showed that, from a stint perspective, there is a clear correlation between optimal stint length and charge time, which corresponds to the maximization of the average stint velocity. Moreover, the optimal race strategy showed a 30\% increase in the overall race distance, compared to a baseline flat-out strategy. 
Finally, we highlighted the importance of optimizing both levels and that, compared to the strategies optimized for a pre-defined number of pit stops, jointly optimizing the number of pit stops can significantly increase the total distance driven by multiple laps, hence considerably improve the achievable race outcome.
%

This work opens the field for the following possible extensions: First, we want to account for the temperature dynamics of the \gls{acr:em} and the battery during driving and charging, since they can play an important role in endurance racing scenarios~\cite{LocatelloKondaEtAl2020,HerrmannPassigatoEtAl2020}.
Second, we want to study the impact of the vehicle dynamics~\cite{HerrmannChristEtAl2019,BroereSalazar2022} and tire degradation on the achievable stint time and the resulting race strategies.



%% file: Kampen.Salazar.ea.ECC22.bbl
\newcommand{\noopsort}[1]{} \newcommand{\printfirst}[2]{#1}
  \newcommand{\singleletter}[1]{#1} \newcommand{\switchargs}[2]{#2#1}
\begin{thebibliography}{10}
\providecommand{\url}[1]{#1}
\csname url@samestyle\endcsname
\providecommand{\newblock}{\relax}
\providecommand{\bibinfo}[2]{#2}
\providecommand{\BIBentrySTDinterwordspacing}{\spaceskip=0pt\relax}
\providecommand{\BIBentryALTinterwordstretchfactor}{4}
\providecommand{\BIBentryALTinterwordspacing}{\spaceskip=\fontdimen2\font plus
\BIBentryALTinterwordstretchfactor\fontdimen3\font minus
  \fontdimen4\font\relax}
\providecommand{\BIBforeignlanguage}[2]{{%
\expandafter\ifx\csname l@#1\endcsname\relax
\typeout{** WARNING: IEEEtran.bst: No hyphenation pattern has been}%
\typeout{** loaded for the language `#1'. Using the pattern for}%
\typeout{** the default language instead.}%
\else
\language=\csname l@#1\endcsname
\fi
#2}}
\providecommand{\BIBdecl}{\relax}
\BIBdecl

\bibitem{FIA2021}
FIA. (2021) 2021 fia world endurance championship: Sporting regulations.
  Available online at
  \url{https://www.fia.com/sites/default/files/2021_wec_sporting_regulations_-_wmsc161220-clean_1.pdf}.

\bibitem{LotEvangelou2013}
R.~Lot and S.~Evangelou, ``Lap time optimization of a sports series hybrid
  electric vehicle,'' in \emph{{World Congress on Engineering}}, 2013.

\bibitem{SedlacekOdenthalEtAl2020}
T.~Sedlacek, D.~Odenthal, and D.~Wollherr, ``Minimum-time optimal control for
  battery electric vehicles with four wheel-independt drives considering
  electrical overloading,'' \emph{{Vehicle System Dynamics}}, 2020.

\bibitem{Casanova2000}
D.~Casanova, ``On minimum time vehicle manoeuvring: The theoretical optimal
  lap,'' Ph.D. dissertation, {Cranfield University}, 2000.

\bibitem{LimebeerPerantoni2014}
D.~Limebeer and G.~Perantoni, ``Optimal control for a formula one car with
  variable parameters,'' \emph{{Vehicle System Dynamics}}, vol.~52, no.~5, pp.
  653--678, 2014.

\bibitem{ChristWischnewskiEtAl2019}
F.~Christ, A.~Wischnewski, A.~Heilmeier, and B.~Lohmann, ``Time-optimal
  trajectory planning for a race car considering variable tyre-road friction
  coefficients,'' \emph{{Vehicle System Dynamics}}, vol.~59, pp. 588--612,
  2019.

\bibitem{DalBiancoLotEtAl2017}
N.~Dal~Bianco, R.~Lot, and M.~Gadola, ``Minimum time optimal control simulation
  of a gp2 race car,'' \emph{{Journal of Automobile Engineering}}, vol. 232,
  pp. 1180--1195, 2017.

\bibitem{HeilmeierWischnewskiEtAl2020}
A.~Heilmeier, A.~Wischnewski, L.~Hermansdorfer, J.~Betz, M.~Lienkamp, and
  B.~Lohmann, ``Minimum curvature trajectory planning and control for an
  autonomous race car,'' \emph{{Vehicle System Dynamics}}, vol.~58, no.~10, pp.
  1497--1527, 2020.

\bibitem{HerrmannChristEtAl2019}
T.~Herrmann, F.~Christ, J.~Betz, and M.~Lienkamp, ``Energy management strategy
  for an autonomous electric racecar using optimal control,'' in \emph{{Proc.\
  IEEE Int.\ Conf.\ on Intelligent Transportation Systems}}, 2019.

\bibitem{HerrmannPassigatoEtAl2020}
T.~Herrmann, F.~Passigato, J.~Betz, and M.~Lienkamp, ``Minimum race-time
  planning-strategy for an autonomous electric racecar,'' in \emph{{Proc.\ IEEE
  Int.\ Conf.\ on Intelligent Transportation Systems}}, 2020.

\bibitem{LiuFotouhiEtAl2020}
X.~Liu, A.~Fotouhi, and D.~Auger, ``Optimal energy management for formula-e
  cars with regulatory limits and thermal constraints,'' \emph{{Applied
  Energy}}, vol. 279, 2020.

\bibitem{HerrmannSauerbeckEtAl2021}
T.~Herrmann, F.~Sauerbeck, M.~Bayerlein, and M.~Betz, J. an~Lienkamp,
  ``Optimization-based real-time-capable energy strategy for autonomous
  electric race cars,'' \emph{{SAE International Journal of Connected and
  Automated Vehicles}}, 2021, in press.

\bibitem{LiuFotouhi2020}
X.~Liu and A.~Fotouhi, ``Formula-e race strategy development using artificial
  neural networks and monte carlo tree search,'' \emph{{Neural Computing and
  Applications}}, no.~32, p. 15191–15207, 2020.

\bibitem{EbbesenSalazarEtAl2018}
S.~Ebbesen, M.~Salazar, P.~Elbert, C.~Bussi, and C.~H. Onder, ``Time-optimal
  control strategies for a hybrid electric race car,'' \emph{{IEEE Transactions
  on Control Systems Technology}}, vol.~26, no.~1, pp. 233--247, 2018.

\bibitem{SalazarElbertEtAl2017}
M.~Salazar, P.~Elbert, S.~Ebbesen, C.~Bussi, and C.~H. Onder, ``Time-optimal
  control policy for a hybrid electric race car,'' \emph{{IEEE Transactions on
  Control Systems Technology}}, vol.~25, no.~6, pp. 1921--1934, 2017.

\bibitem{DuhrChristodoulouEtAl2020}
P.~Duhr, G.~Christodoulou, C.~Balerna, M.~Salazar, A.~Cerofolini, and C.~H.
  Onder, ``Time-optimal gearshift and energy management strategies for a hybrid
  electric race car,'' \emph{{Applied Energy}}, vol. 282, no. 115980, 2020.

\bibitem{BorsboomFahdzyanaEtAl2021}
O.~Borsboom, C.~A. Fahdzyana, T.~Hofman, and M.~Salazar, ``A convex
  optimization framework for minimum lap time design and control of electric
  race cars,'' \emph{{IEEE Transactions on Vehicular Technology}}, vol.~70,
  no.~9, pp. 8478--8489, 2021.

\bibitem{LocatelloKondaEtAl2020}
A.~Locatello, M.~Konda, O.~Borsboom, T.~Hofman, and M.~Salazar, ``Time-optimal
  control of electric race cars under thermal constraints,'' in \emph{{European
  Control Conference}}, 2021.

\bibitem{HeilmeierGrafEtAl2018}
A.~Heilmeier, M.~Graf, and M.~Lienkamp, ``A race simulation for strategy
  decisions and circuit motorsports,'' in \emph{{Proc.\ IEEE Int.\ Conf.\ on
  Intelligent Transportation Systems}}, 2018.

\bibitem{WestLimebeer2020}
W.~J. West and D.~J.~N. Limebeer, ``Optimal tyre management for a
  high-performance race car,'' \emph{{Vehicle System Dynamics}}, vol. 231, pp.
  1--19, 2020.

\bibitem{GuzzellaSciarretta2007}
L.~Guzzella and A.~Sciarretta, \emph{Vehicle propulsion systems: Introduction
  to Modeling and Optimization}, 2nd~ed.\hskip 1em plus 0.5em minus 0.4em\relax
  {Springer Berlin Heidelberg}, 2007.

\bibitem{BorsboomFahdzyanaEtAl2020}
O.~Borsboom, C.~A. Fahdzyana, M.~Salazar, and T.~Hofman, ``Time-optimal control
  strategies for electric race cars for different transmission technologies,''
  in \emph{{IEEE Vehicle Power and Propulsion Conference}}, 2020.

\bibitem{BoydVandenberghe2004}
S.~Boyd and L.~Vandenberghe, \emph{Convex optimization}.\hskip 1em plus 0.5em
  minus 0.4em\relax {Cambridge Univ.\ Press}, 2004.

\bibitem{RichardsHow2005}
A.~Richards and J.~How, ``Mixed-integer programming for control,'' in
  \emph{{American Control Conference}}, 2005.

\bibitem{InMotion}
\BIBentryALTinterwordspacing
``{InMotion fully electric LMP3 car}.'' [Online]. Available:
  \url{https://www.inmotion.tue.nl/en/about-us/cars/new-car}
\BIBentrySTDinterwordspacing

\bibitem{AnderssonGillisEtAl2019}
J.~A.~E. Andersson, J.~Gillis, G.~Horn, J.~B. Rawlings, and M.~Diehl, ``Casadi
  -- a software framework for nonlinear optimization and optimal control,''
  \emph{{Mathematical Programming Computation}}, vol.~11, no.~1, pp. 1--36,
  2019.

\bibitem{WachterBiegler2006}
A.~Wachter and L.~T. Biegler, ``On the implementation of an interior-point
  filter line-search algorithm for large-scale nonlinear programming,''
  \emph{{Mathematical Programming}}, vol. 106, no.~1, pp. 25--57, 2006.

\bibitem{HSL}
\BIBentryALTinterwordspacing
HSL. A collection of fortran codes for large scale scientific computation.
  [Online]. Available: \url{http://www.hsl.rl.ac.uk}
\BIBentrySTDinterwordspacing

\bibitem{Loefberg2004}
J.~L{\"o}fberg, ``{YALMIP} : A toolbox for modeling and optimization in
  {MATLAB},'' in \emph{{IEEE Int.\ Symp.\ on Computer Aided Control Systems
  Design}}, 2004.

\bibitem{ApS2017}
M.~{ApS}. (2017) {MOSEK} optimization software. {Available at
  }\url{https://mosek.com/}.

\bibitem{BroereSalazar2022}
S.~Broere and M.~Salazar, ``Minimum-lap-time control strategies for all-wheel
  drive electric race cars via convex optimization,'' in \emph{{European
  Control Conference}}, 2022, in press. Available online at
  \url{https://arxiv.org/abs/2111.04650}.

\end{thebibliography}
